\documentclass[traditabstract,print]{aa}

\usepackage{graphicx}
\usepackage{txfonts}
\usepackage{multirow}
\usepackage{fancyhdr} 
\usepackage{enumerate}
\usepackage[authoryear]{natbib}
\usepackage{longtable} 
\usepackage{subfigure}
\usepackage{deluxetable}
\begin{document}
\title{ The North Ecliptic Pole Wide survey of AKARI: a near- and mid-infrared source catalog
\footnote{The full version of Table 4 is only available in electronic form at the CDS via anonymous ftp to
cdsarc.u-strasbg.fr (130.79.128.5) or via http://cdsarc.u-strasbg.fr/cgi-bin/qcat?J/A+A/ } }
\author{Seong Jin Kim\inst{1},
        Hyung Mok Lee\inst{1},
        Hideo Matsuhara\inst{2},
        Takehiko Wada\inst{2},
        Shinki Oyabu\inst{3},
        Myungshin Im\inst{1},
        Yiseul Jeon\inst{1},
        Eugene Kang\inst{4},
        Jongwan Ko\inst{4,5},
        Myung Gyoon Lee\inst{1}
        Toshinobu Takagi\inst{2},
        Chris Pearson\inst{6},
        Glenn J. White\inst{6, 7},
        Woong-Seob Jeong\inst{4},
        Stephen Serjeant\inst{7},
        Takao Nakagawa\inst{2},
        Youichi Ohyama\inst{8},
        Tomotsugu Goto\inst{9},
        Tsutomu T. Takeuchi\inst{10},
        Agnieszka Pollo\inst{11, 12, 13},
        Aleksandra Solarz\inst{10},
        Agata P\c{e}piak\inst{11} }

\institute{Astronomy Program, Department of Physics and Astronomy, FPRD, 
          Seoul National University, Kwanak-Gu, Seoul 151-742, Republic of Korea \\
           \email{seongini@astro.snu.ac.kr}
\and
Institute of Space and Astronautical Science, Japan Aerospace 
          Exploration Agency, Sagamihara, 229-8510 Kanagawa, Japan
\and
Graduate School of Science, Nagoya University, Furo-cho, Chikusa-ku, 
          Nagoya, 464-8602 Aichi, Japan
\and
Korea Astronomy and Space Science Institute, Deajeon 305-348, Republic of Korea
\and        
Yonsei University Observatory, Yonsei University, Seoul 120-749, Republic of Korea
\and 
RALSpace, The Rutherford Appleton Laboratory, Chilton, Didcot, xfordshire OX11 0QX, UK
\and
Astrophysics Group, Department of Physics, The Open University, Milton Keynes, MK7 6AA, UK
\and
Academia Sinica, Institute of Astronomy and Astrophysics, Taiwan
\and
Institute for Astronomy, University of Hawaii, 2680 Woodlawn Drive, Honolulu, HI, 96822, USA
\and
Department of Particle and Astrophysical Science, Nagoya University, Furo-cho, 
 Chikusa-ku, Nagoya 464-8602, JAPAN 
\and
The Astronomical Observatory of the Jagiellonian University, ul.  Orla 171, 30-244 Krak\'{o}w, POLAND
\and
Polish Academy of Sciences, al.  Lotnik\'{o}w, 32/46, 02-668, Warsaw, POLAND
\and
The Andrzej So{\l}tan National Centre for Nuclear Research, ul. Ho\.{z}a 69, 00-681 Warsaw, POLAND
}

\authorrunning{S. J. Kim et al.}
\titlerunning{The AKARI NEP-Wide survey : a point source catalog}

\date {Received February 24, 2012; accepted July 24, 2012}

\abstract
{We present a photometric catalog of infrared (IR) sources based on the North Ecliptic Pole Wide 
field (NEP-Wide) survey of AKARI, which is an infrared space telescope launched by Japan. The NEP-Wide 
survey covered 5.4 deg$^{2}$ area, a nearly circular shape centered on the North Ecliptic Pole, using 
nine photometric filter-bands from 2 -- 25 $\mu$m of the Infrared Camera (IRC). Extensive efforts were 
made to reduce possible false objects due to cosmic ray hits, multiplexer bleeding phenomena around 
bright sources, and other artifacts. The number of detected sources varied depending on the filter 
band: with about 109,000 sources being cataloged in the near-IR bands at 2 -- 5 $\mu$m, about 20,000 
sources in the shorter parts of the mid-IR bands between 7 -- 11 $\mu$m, and about 16,000 
sources in the longer parts of the mid-IR bands, with  $\sim$ 4,000 sources at 24 $\mu$m. The 
estimated 5$\sigma$ detection limits are approximately 21 magnitude (mag) in the 2 -- 5 $\mu$m bands, 
19.5 -- 19 mag in the 7 -- 11 $\mu$m, and 18.8 -- 18.5 mag in the 15 -- 24 $\mu$m bands in the AB 
magnitude scale. The completenesses for those bands were evaluated as a function of magnitude: the 
50$\%$ completeness limits are about 19.8 mag at 3 $\mu$m, 18.6 mag at 9 $\mu$m, and 18 mag at 18 $\mu$m 
band, respectively. To construct a reliable source catalog, all of the detected sources were examined 
by matching them with those in other wavelength data, including optical and ground-based near-IR bands.
The final band-merged catalog contains about {114,800} sources  detected in the IRC filter bands. The 
properties of the sources are presented in terms of the distributions in various color-color diagrams.
}

\maketitle

\section{INTRODUCTION}

 An infrared (IR) space telescope AKARI, launched by ISAS/JAXA in 2006 (Murakami et al. 2007), 
successfully carried out an all-sky survey at mid- and far-infrared wavelengths, as well as several 
large-area surveys and many other pointed observations across the wavelength range 2 -- 160 $\mu$m. 
The North Ecliptic Pole (NEP) survey (Matsuhara et al. 2006) was one of these large area surveys of 
AKARI. The AKARI telescope had two focal-plane instruments: the Infrared Camera (IRC, Onaka et al. 2007) 
and the Far-Infrared Surveyor (FIS, Kawada et al. 2007). The FIS covered a 50 -- 200 $\mu$m range with
four wide-band photometric filters, and a Fourier Transform Spectrometer (FTS). The IRC was designed to 
carry out near- to mid-infrared imaging with nine photometric filters, and spectroscopic observations 
with a prism and grisms.

\begin{figure}[h]
\begin{center}
\resizebox{\hsize}{!}{\includegraphics{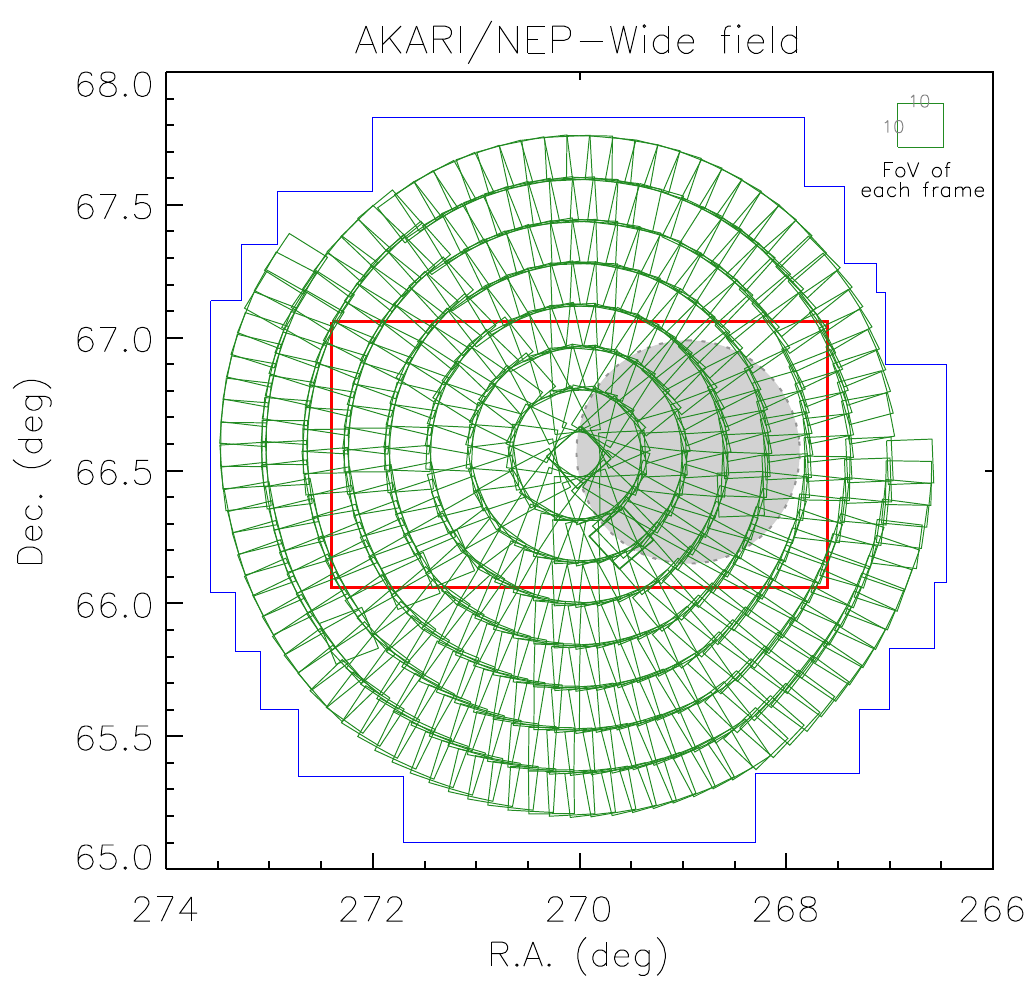}}
\caption{The overall map of the NEP-Wide field. The survey consisted of 446 pointing observations
represented by green boxes. Each frame covered a 10$'\times$10$'$ area with half of its field of view
(FoV) overlapped by neighboring frames. The red box and blue lines represent the regions covered by
optical surveys at the CFHT and  Maidanak Observatory, respectively. The circular gray-shaded region
denotes the NEP-Deep field.}
\label{fig 01}
\end{center}
\end{figure}

The NEP survey is composed of two parts: a wide (henceforth NEP-Wide) survey and a deep (henceforth 
NEP-Deep) survey. The area observed in the NEP-Wide survey is shown in Fig. 1 (green tiles) and is 
about 5.4 deg$^{2}$ with a circular shape (whose radius is about 1.25 deg) centered on the NEP 
($\alpha =18^{h}00^{m}00^{s}$, $\delta= +66^{\circ}33^{'}38^{''} $), while the NEP-Deep survey covers 
about 0.6 deg$^{2}$ (Wada et al. 2008, Takagi et al. 2012), with a center slightly offset from the NEP 
(shaded region).  These surveys were observed using the nine IRC bands to provide nearly continuous 
coverage from 2 $\mu$m to 25$\mu$m.  The filter system is designated as $N2$, $N3$, and $N4$ for the 
near-infrared (NIR) bands, $S7$, $S9W$, and $S11$ for the shorter part of the mid-IR band (MIR-S), 
and $L15$, $L18W$, and $L24$ for the longer part of the mid-IR bands (MIR-L) with the numbers 
representing the approximate effective wavelengths in units of $\mu$m. The photometric bands with 
wider spectral widths are indicated by a W at the end. The NEP-Wide survey was carried out with 446 
pointed observations, with the field of view (FoV) of each individual frame being 10$'\times$10$'$. 
The survey coverage was designed around seven concentric circles and a partial rim in the outermost 
part.

The details of the strategy, the observational plans for the coordinated pointing surveys, the 
scientific goals and the technical constraints are described in Matsuhara et al. (2006). The initial 
results and the catalog for the NEP-Deep survey were reported by Wada et al. (2008). The data 
characteristics and basic properties of the sources were presented using a subset of NEP-Wide data 
by Lee et al. (2009). Here, we present the entire data set of NEP-Wide, corrected to remove artificial 
effects, and provide a more detailed analysis of the photometric results. The main purposes of this 
paper is to present the data reduction methodology, and provide a point source catalog of the NEP-Wide 
survey. 

This paper is organized as follows. In sect. 2, we present details of the data reduction process, 
and the additional image corrections needed to improve the efficacy of the image data.  We describe 
the results of our source extraction and photometry, as well as the properties of the data such as 
sensitivity and the completeness of  the source detection in \S 3.  The next section describes 
the source matching across the available bands to confirm the genuineness of the detected sources 
in \S 4. In \S 5, we present the band-merging procedure and describe the contents in the final 
catalog. The natures of the detected sources using various color-magnitude diagrams and color-color 
diagrams are shown in \S 6.  We summarize  our results in the final section.

\section{DATA REDUCTION }

\subsection{Pre-processing }

\subsubsection{Standard reduction with IRC pipeline }

The individual pointing data from the AKARI NEP-Wide field was obtained using the observation
template `IRC03' (Onaka et al. 2007), which was designed for general imaging observations. During a 
single pointing observation with IRC03, each exposure consisted of observations with three combined 
filters (two mid-IR bands and one near-IR band).  Fig. 2 shows the overall view of the observation 
sequence, as well as the structure of the raw data obtained from a single pointing observation using 
`IRC03'. Each pointing data was reduced by the IRC imaging pipeline (Lorente et al. 2008) implemented 
in the IRAF\footnote{IRAF is distributed by the National Optical Astronomy Observatories, which are 
operated by the Association of Universities for Research in Astronomy, Inc., under cooperative 
agreement with the National Science Foundation.} environment. We adopted the official package version 
071017\footnote{This software is accessible at AKARI observers web page 
(http://www.ir.isas.jaxa.jp/AKARI/Observation).} without any modification of the CL script. The 
pipeline is composed of three stages, correcting for the instrumental features and transforming 
the raw data packets to basic science data as described below.

\begin{figure}[h]
\begin{center}
\resizebox{\hsize}{!}{\includegraphics{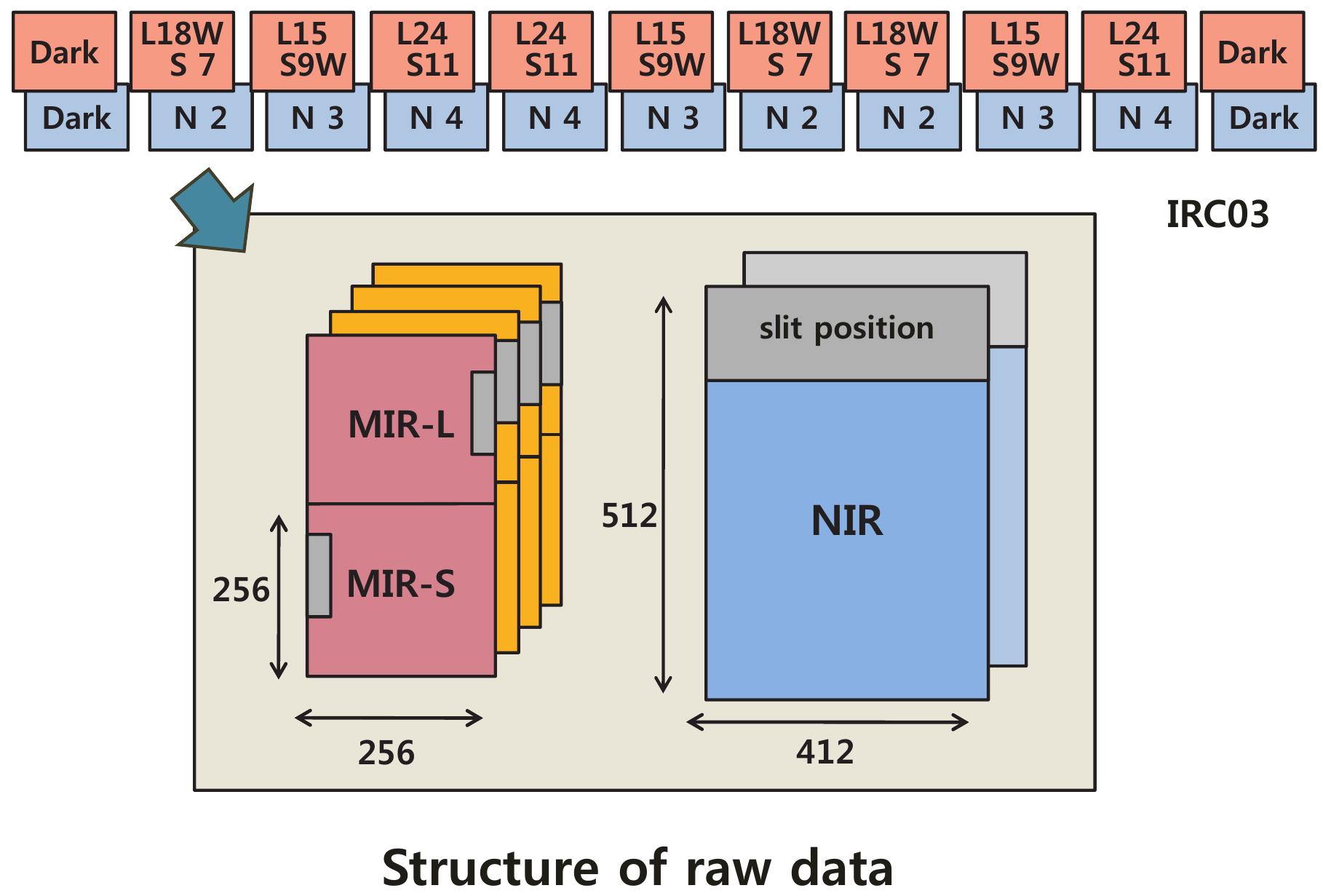}}
\caption{An overall view of the observation template IRC03, which was used for NEP-Wide observations.
For one pointing observation, the exposure was carried out according to a fixed sequence. The structure
of the raw data obtained in the IRC03 mode are shown in the bottom panel. The NIR data consist of one
short and one long-exposure, whereas the MIR data are composed of one short- and three long-exposures.
\label{fig 2}}
\end{center}
\end{figure}

The conceptual structure of the pipeline is schematically shown in Fig. 3. The structure of the
packaged raw data for each individual pointed observation is a three-dimensional cube that consists 
of combined single frames created during each exposure cycle as shown in Fig. 2.  The first stage, 
called the `red-box' module, slices these into standard two-dimensional image frames and creates 
an observation log containing the summary of the processed files in the working directory.  Each 
data set can be recognized by its target name, filter name, pointing identification number (PID), 
and coordinates (R.A., Dec.).  Most of these procedures are automatically carried out by running 
the `prepipeline' command.

\begin{figure}[h]
\begin{center}
\resizebox{\hsize}{!}{\includegraphics{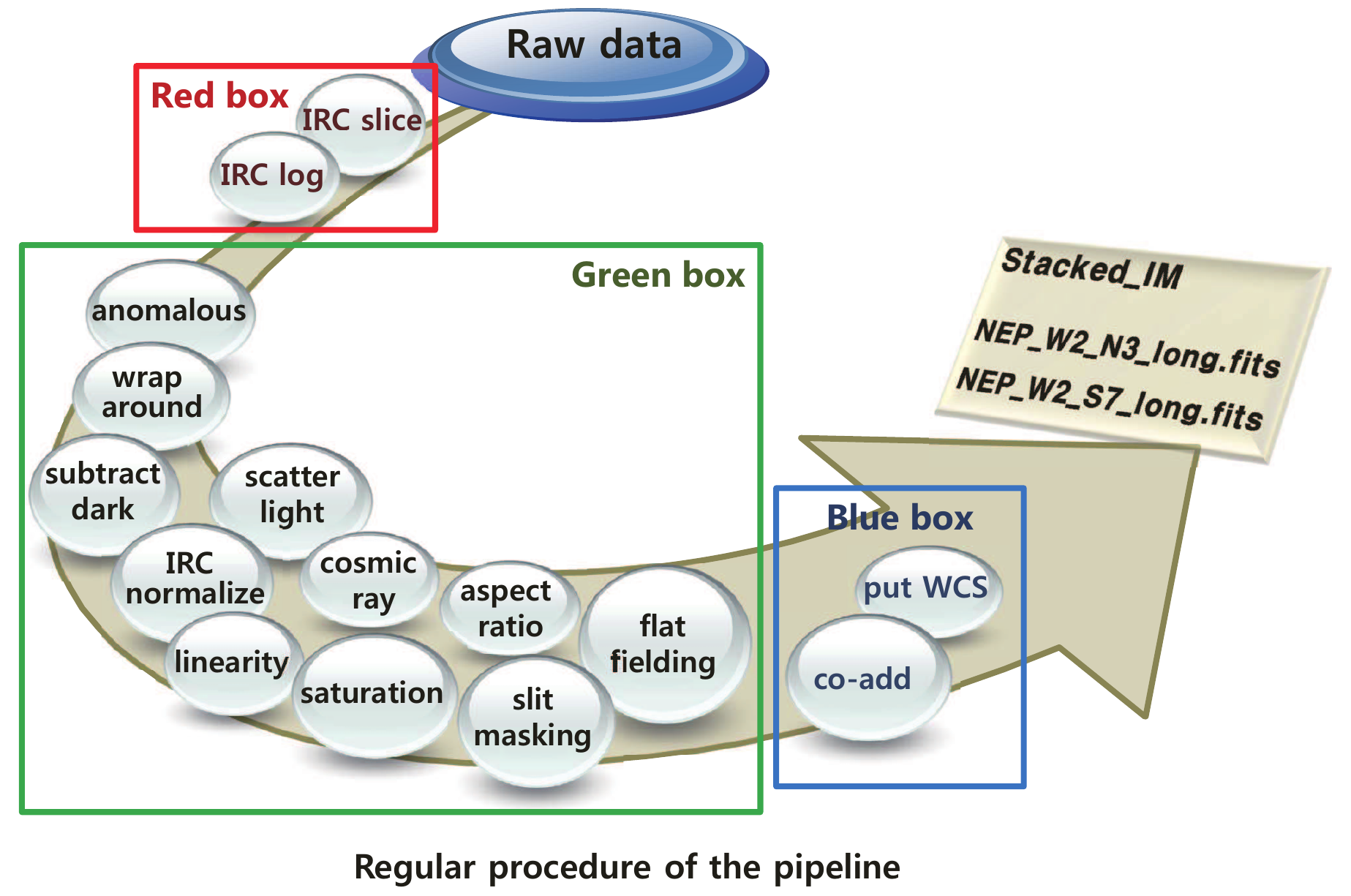}}
\caption{A schematic view of the procedures in the pipeline designed for the preprocessing
of the IRC imaging data.  The imaging pipeline is composed of three parts.
\label{fig 3.}}
\end{center}
\end{figure}

In the second stage (the green-box module), the pipeline performs various procedures such as
masking of bad pixels, subtraction of dark current, linearization of detector response, and 
correction for distortion and flat fielding. We chose a `self-dark' parameter and the pipeline's 
default flat. The self-dark is an image made by averaging pre-dark images of MIR-S and MIR-L 
long-exposure dark frames from each pointed observation (Lorente et al. 2008). Here, the pre-dark 
is the dark frame taken at the beginning of the operation for the NIR and MIR channels (Fig. 2). 
The super-dark was obtained from more than 100 pointings of pre-dark images taken at the early 
stage of the mission to provide superior signal to noise (S/N). However, self-dark may be able to 
remove hot pixels more efficiently than using the super-dark, especially for the  MIR-S and MIR-L 
data. Hence, for the NIR long-exposure frames, the super-dark is used to achieve higher S/N. In 
the MIR data, the cosmic ray hits were removed at this stage. However, for 19 frames of the $S9W$ 
and 104 frames of $S11$, there remains a small noticeable pattern in the lower right part even 
after flat-fielding.  We simply removed the area having bright patterns after the pre-processing 
stage.

The third stage (the blue-box) calculates the relative shifts and rotations among the frames
before stacking, in order to match the attitude of the frames, since the individual frames taken 
at a given pointing observation are not exactly aligned.  In addition, the blue-box stage estimates 
the average sky and adjusts the sky levels before stacking these frames. We chose the default option 
`submedsky' for those processes.  At this point, to stack the MIR-L frames, we used `coaddLusingS'. 
This is an optional task in the IRC pipeline, that utilizes the stacking information from the MIR-S 
frames, which are simultaneously obtained with MIR-L frames, thus have identical PIDs. We used this 
method because a successful coaddition is guaranteed thanks to the same rotations and shifts as the 
MIR-S frames of the same PID, even though some of them are not properly stacked when using only 
MIR-L images. For these tasks, we used long-exposure frames, and selected 3$\sigma$ for the limits 
of image statistics in the pipeline reduction.  Finally, the IRC pipeline adds header information 
that supports the world coordinate system (WCS).

\subsubsection{Astrometry }

 To input the information of the WCS into the stacked frame, the toolkit `putwcs' of the pipeline 
calculates astrometry by matching the bright point sources in the IRC image to reference objects 
using the Two-Micron All-Sky Survey (2MASS) data (Skrutskie et al.  2006). The astrometric accuracy 
of the `putwcs' is less than $1^{\prime \prime}$ rms for the NIR bands, about $2^{\prime \prime}$ 
rms for MIR-S bands, and $3^{\prime \prime}$ -- $4^{\prime \prime}$ rms for MIR-L bands (Wada et al. 
2008; Shim et al. 2011). The full-width-at-half-maximum (FWHM) of stacked frames from a pointing 
observation are about $4^{\prime \prime}.2$ in the NIR bands, and in the range $5^{\prime \prime}.2$ 
-- $6^{\prime \prime}.8$ in the MIR bands (Lorente et al. 2008; Lee et al. 2009) depending on the 
filter.

The `putwcs' was run automatically from the $N2$ to $S9W$ band images. However, this task did not work 
well for the wavelength bands longer than $S9W$ because of an insufficient number of sources with 
2MASS counterparts. For this reason, the astrometry for about 40 frames ($\sim 9\%$) of the $S11$ 
band was derived by comparing with the $S9W$ frames whose WCS solutions had already been determined.  
This alternative method was satisfactory because the $S11$ frames with unsuccessful `putwcs' 
operations constituted only a small fraction of the total, and the sources in these $S11$ frames 
were easily identified using $S9W$ data with the same PID. At this stage, three sets of pointing 
data from the NIR bands were discarded owing to stacking failures and very low quality (PID : 
2110888). For the $L15$ band, the automatic operation of `putwcs' was successful for only 84$\%$ 
frames. The astrometry of the remaining $\sim 16\%$ was done using $S11$ sources in the same manner 
as described above for $S11$ frames. However, the pointing direction of the MIR-L channels is $\sim$ 
20.6$'$ apart from that of the MIR-S, while the NIR and the MIR-S share the same field of view. 
Therefore, to use $S11$ sources for the astrometry of the $L15$ data, we used pertinent regions 
extracted from the mosaicked image of the $S11$ band.

\begin{figure}[h]
\begin{center}
\resizebox{\hsize}{!}{\includegraphics{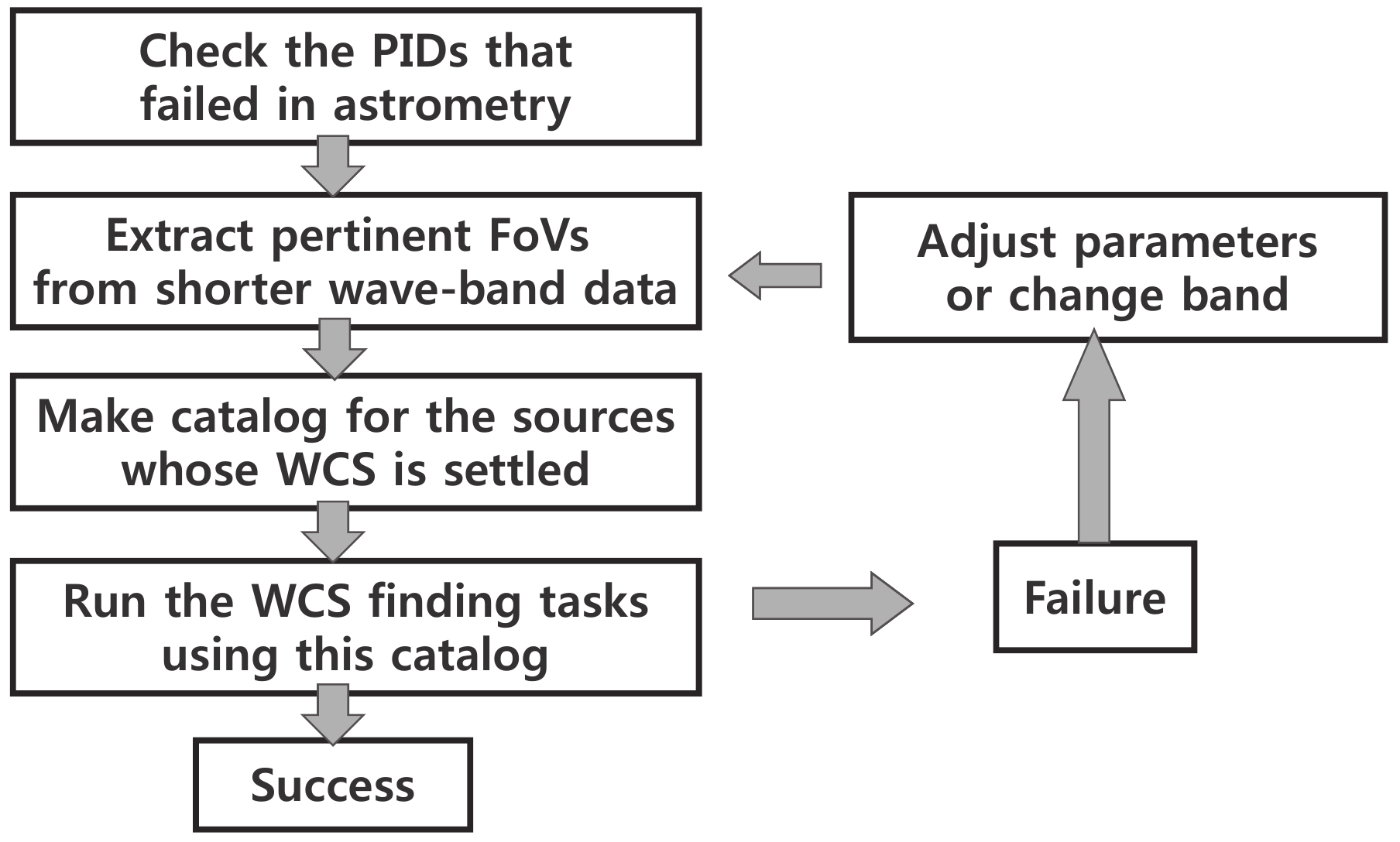}}
\caption{Schematic flow chart describing the iterative procedures used to find a satisfactory
solution for the WCS header of the $L18W$ and $L24$ band data.
\label{fig 4.}}
\end{center}
\end{figure}

The astrometric solutions for the $L18W$ data were obtained in a similar way to that for the $L15$ 
band. However, the number of common sources between the $L18W$ band and 2MASS data is quite small 
so that the `putwcs' operations were applicable to only about 50$\%$ of the $L18W$ frames. The 
remaining frames were processed using the $L15$ sources, but these attempts were also not fully 
successful because, for many frames, the number of $L18W$ sources cross-matched with $L15$ sources 
was insufficient. For these frames, we used the corresponding region extracted from the mosaicked 
images of the $S11$ and $S9W$ bands. For the $L24$ band, the automatic execution of `putwcs' was 
able to find the astrometric solution for only $\sim$ 10$\%$ frames of the total since 24 $\mu$m 
sources with 2MASS counterparts are very rare. We therefore conducted an alternative source-matching 
with the $L18W$ and $L15$ band data and obtained the astrometry for about 65$\%$ of the frames.  
To get correct astrometry using WCS deriving tasks, we have to ensure a sufficient number of sources 
effectively available for the fitting process of the software. The iterative procedures adopted to 
find a sufficient number of sources for the astrometry of the $L18W$ and $L24$ data are shown 
schematically in Fig. 4.  We were able to obtain astrometric solutions for about 90\% of the $L24$ 
frames, leaving $\sim$ 10\% of the frames unusable for source detection and photometry.

\subsection{Post-processing and image correction}

\subsubsection{ Cosmic ray rejection in the NIR bands }

The `IRC03' template takes three long-exposure frames for each NIR band. The last frame $N4$ in 
each pointed observation may be taken during a satellite maneuver to the next pointing, and can be 
automatically rejected by the pipeline when the data quality is too poor (Fig. 2). For this reason, 
most of the $N4$ pointing data have two frames stacked by the pipeline. This causes difficulties 
in removing cosmic rays from $N4$ frames (as well as a small fraction of the $N2$ and $N3$ frames).
Therefore, an alternative method to remove cosmic rays from individual frames has to be applied to 
those frames before the mosaicking. We used a program `L.A. cosmic' which is based on the Laplacian 
edge detection algorithm for highlighting boundaries (van Dokkum 2001). We used the imaging 
version\footnote{There are imaging version and spectroscopic version for IRAF users.} 
of the software written in CL script for IRAF users provided by van Dokkum\footnote{See 
http://www.astro.yale.edu/dokkum/lacosmic}. This procedure relies on the sharpness of the edges 
rather than the contrast between entire cosmic rays and their surroundings, therefore it is 
independent of their shapes. The L.A. cosmic procedure was run by the implementation in IRAF with 
the default settings. This algorithm was found to be quite robust and it effectively rejected 
remaining cosmic rays of arbitrary size.

\begin{figure}[h!]
\begin{center}
\resizebox{\hsize}{!}{\includegraphics{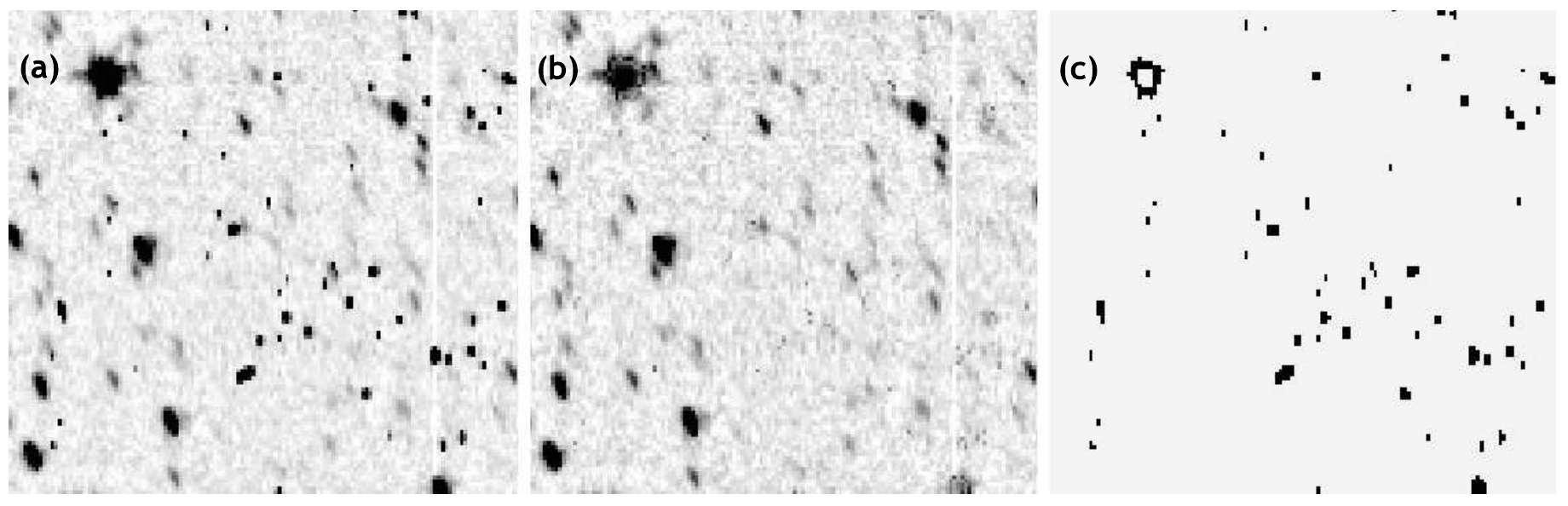}}
\caption{An example showing the result of cosmic-ray rejection. The left panel (a) is a sample 
portion of the $N4$ image produced by the IRC imaging pipeline (PID : 2100757) on which many 
cosmic rays remain. The middle panel (b) is the same image restored by cosmic-ray rejection using 
the L.A.  cosmic procedure.  The third panel (c) shows the result of subtraction,  (a) - (b), 
which is just a map showing the rejected cosmic rays.}
\label{fig 5.}
\end{center}
\end{figure}

Fig. 5 shows an example comparison of the images before (left panel) and after the procedure (middle 
panel). The right panel shows the difference between the two images. However, the sources directly 
hit by cosmic rays are consequently deformed or damaged during the L.A. cosmic procedure, hence the 
photometry for those sources are inevitably affected. Most of these sources damaged by cosmic rays 
were finally rejected by the masking process during the image correction to remove MUX-bleeding 
effects in the NIR bands, as described in the following section.

\subsubsection{Correction for MUX-bleeding effects}

\begin{figure}[ht!]
\begin{center}
\resizebox{\hsize}{!}{\includegraphics{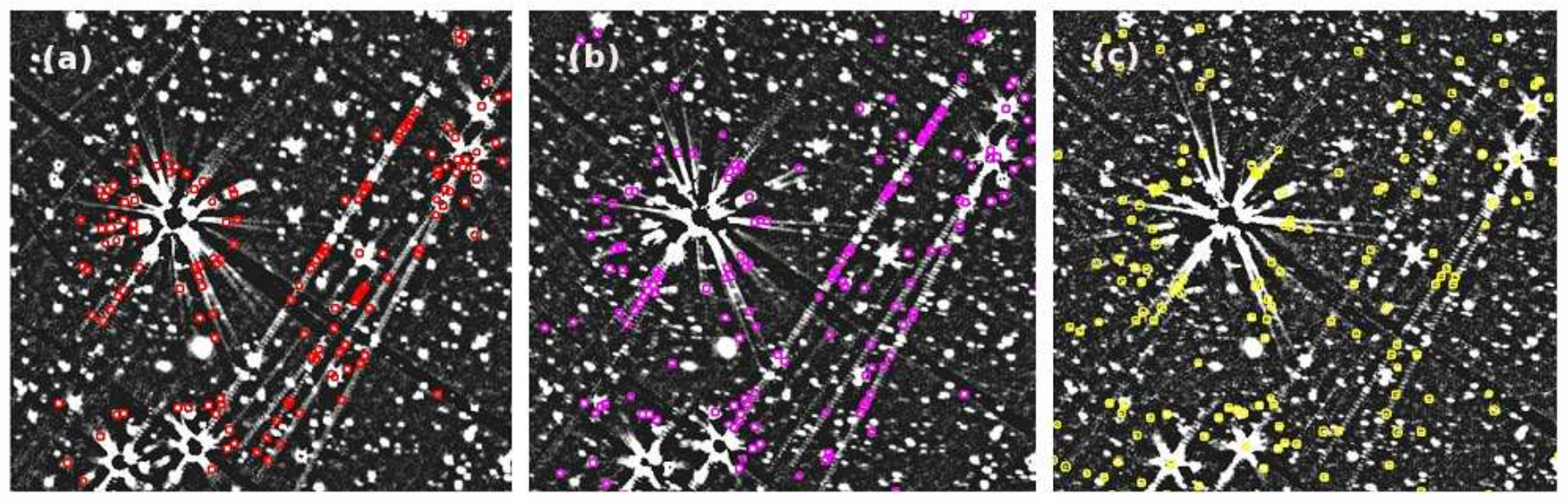}}
\caption{Close-up views of a sample region around bright sources showing false detections at each
NIR band caused by MUX-bleeding trails before the correction. The left, middle, and right panels
show the $N2$, $N3$, and $N4$ band images, respectively. Colored circles indicate the objects that
were not matched with optical data.}
\label{fig 6.}
\end{center}
\end{figure}

\begin{figure*}[ht!]
\centering
\resizebox{\hsize}{!}{\includegraphics{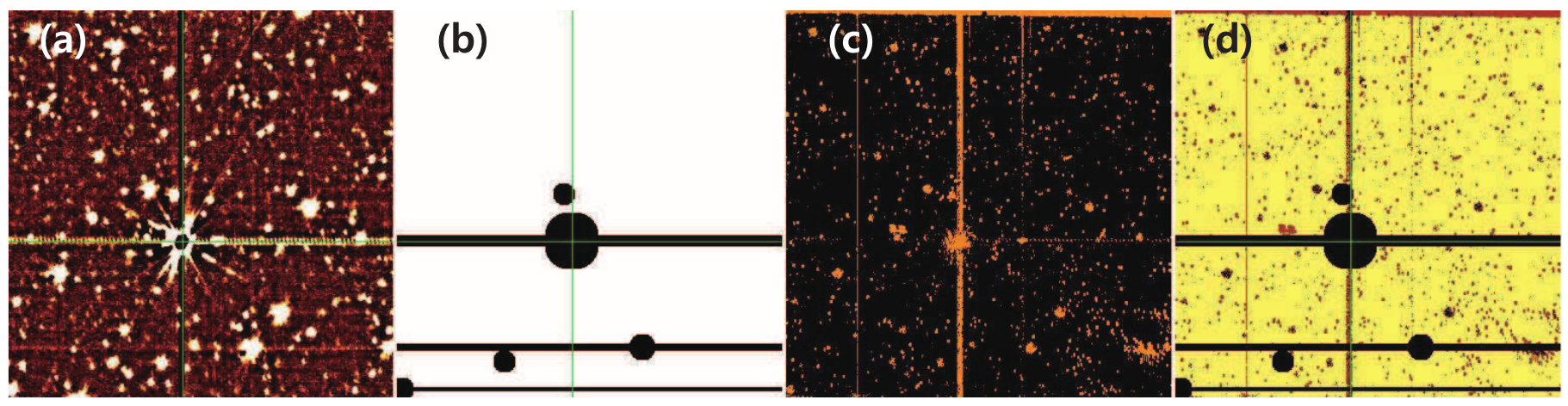}}
\caption{ Schematic diagram describing the procedures used to make a weight map for an individual
frame obtained from a single pointing observation. The first panel (a) is a sample image containing
a bleeding trail and a bright source whose center is disrupted. The second panel (b) shows the regions
to be masked. The third panel (c) represents the number of rejected frames during the stacking
procedure. Each pixel has an integer value ranging 0 to 3. The last panel (d) is a weight map to
be used for the final mosaicking procedure.}
\label{fig 7}
\end{figure*}

In addition to cosmic rays, multiplexer bleed trails (or MUXbleeds, hereafter) remained along the 
horizontal direction in the NIR data because of the nature of InSb detector array (Holloway 1986; 
Offenberg 2001). In general, when there is a bright source in a frame, periodic horizontal features 
appear along the same row. Consequently, many spurious artifacts are detected as sources along the 
bleeding trails.  Fig. 6 shows the region where the spurious detection is serious due to MUXbleeds. 
To mitigate such artifacts, and  facilitate detection and accurate positional matching with real 
source in the other bands, we have to apply an appropriate rejection procedure. However, it is 
difficult to accurately correct for the bleeding effect without any influence on the real sources.  
We simply masked the regions of MUXbleeds, at the expense of rejecting a small number of real 
sources.

The method we used to mask the region severely affected by bleeding effects  assigns a different 
weight to the selected regions when we create a mosaic image with the software SWarp\footnote{See
http://terapix.iap.fr/IMG/pdf/swarp.pdf}. When we wished to remove a region affected by MUXbleeds 
we gave a zero weight to that region. Fig. 7 shows a schematic overview  describing  the steps
adopted to determine the region to be masked and to produce a weight map, which is used to generate 
a mosaic image at the final stage. The leftmost panel (a) is a sample image reduced by the IRC 
pipeline. In the second panel (b), dark stripes and circles show the regions to be masked by 
assigning a zero weight, whereas the white regions have 100$\%$ weight. The third image (c) shows 
the number of rejected frames during the stacking procedure in the final stage of the pipeline. 
For example, there are three dithered exposures in the $N2$ band and each pixel in this map may 
have an integer value ranging from 0 to 3. If we multiply this  by `mask image' of the second 
panel, we can make the final weight map in the panel (d), which  gives proper weight factors for 
individual pixels. Here, the most important task is how to decide the region to be masked to 
effectively remove the muxbleeds while minimizing the pixel losses that were not affected by 
the bleeding trail.

\begin{figure*}[ht!]
\begin{center}
\resizebox{\hsize}{!}{\includegraphics{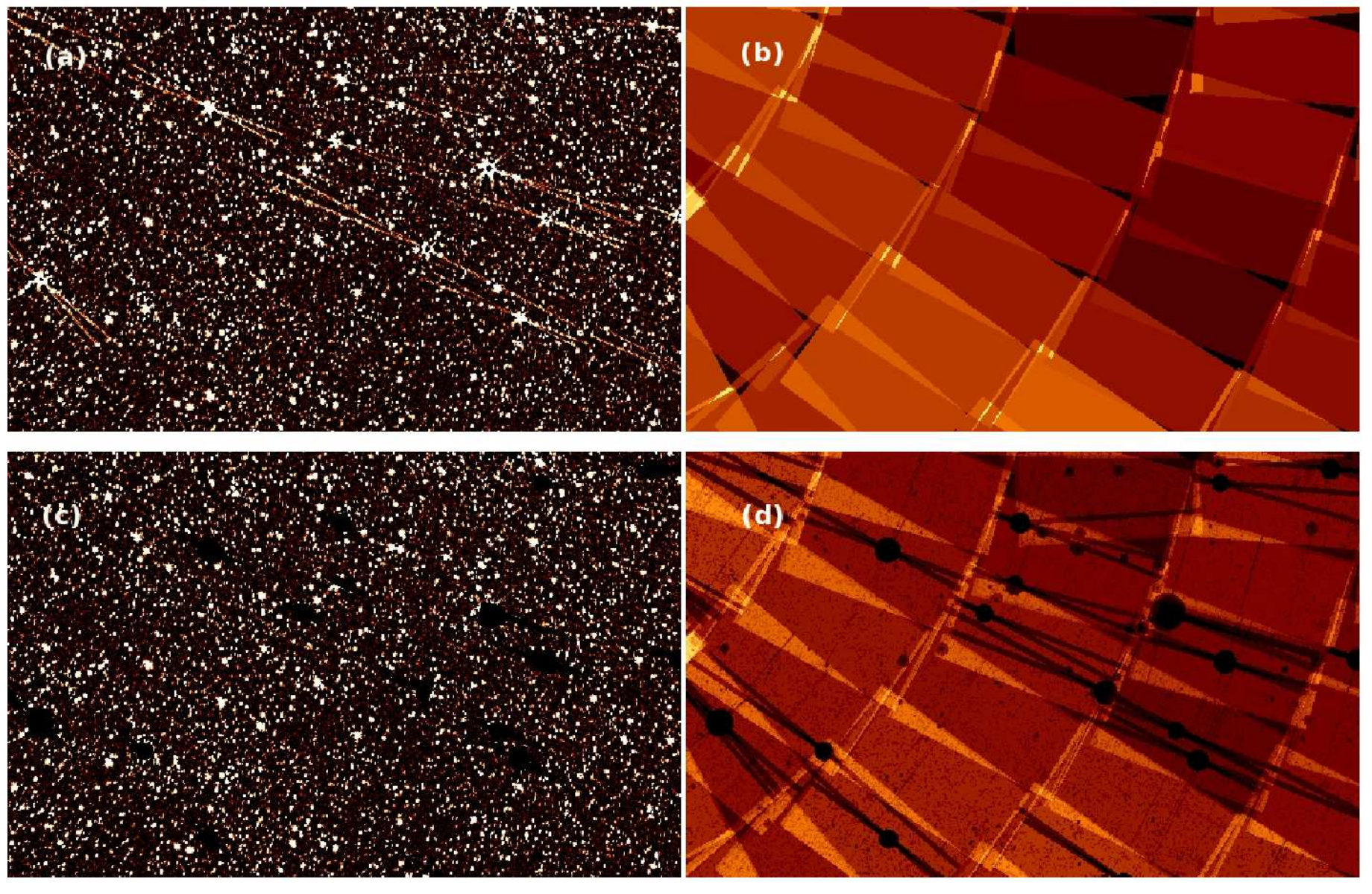}}
\caption{Segments of mosaic images (left) and weighted coverage maps (right)  before
and after the MUXbleed correction ($N3$). The upper pair shows the images before this process and
the bottom panels show them after the operation. The bleeding trails in this field are completely
removed by this weighted mosaicking method using weight maps. In the map (d), zero- weighted pixels
are in black.}
\label{fig 8}
\end{center}
\end{figure*}

\begin{figure*}[ht!]
\begin{center}
\resizebox{0.95\hsize}{!}{\includegraphics{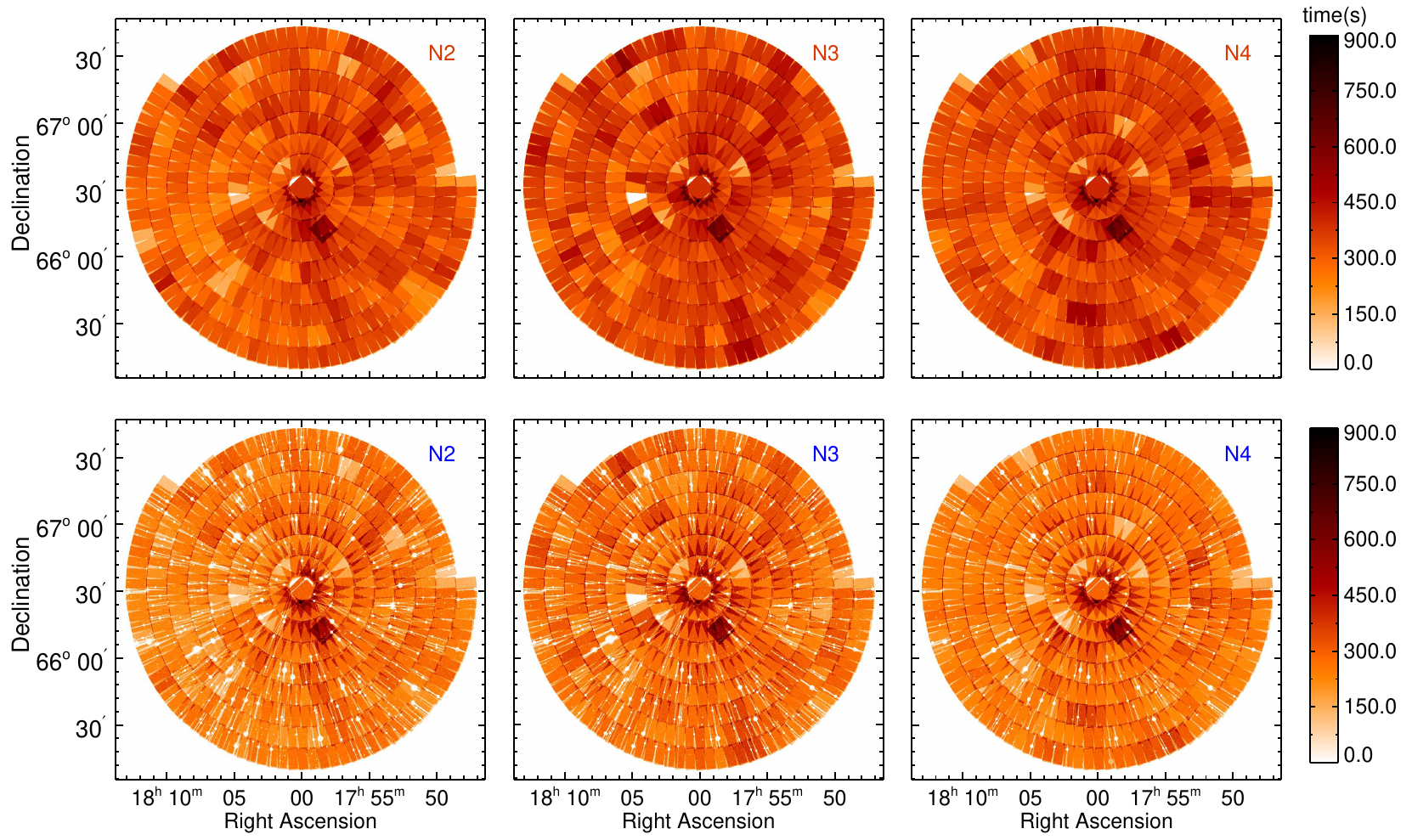}}
\caption{Entire coverage maps of the NIR data  before (upper) and after (lower) the
correction for MUX bleeds. Uncovered area and none-weighted pixels have zero values
(white area), which have wedge-like shape from the center to radially outer direction.
\label{fig 9}}
\end{center}
\end{figure*}

\begin{figure*}[ht!]
\begin{center}
\resizebox{0.95\hsize}{!}{\includegraphics{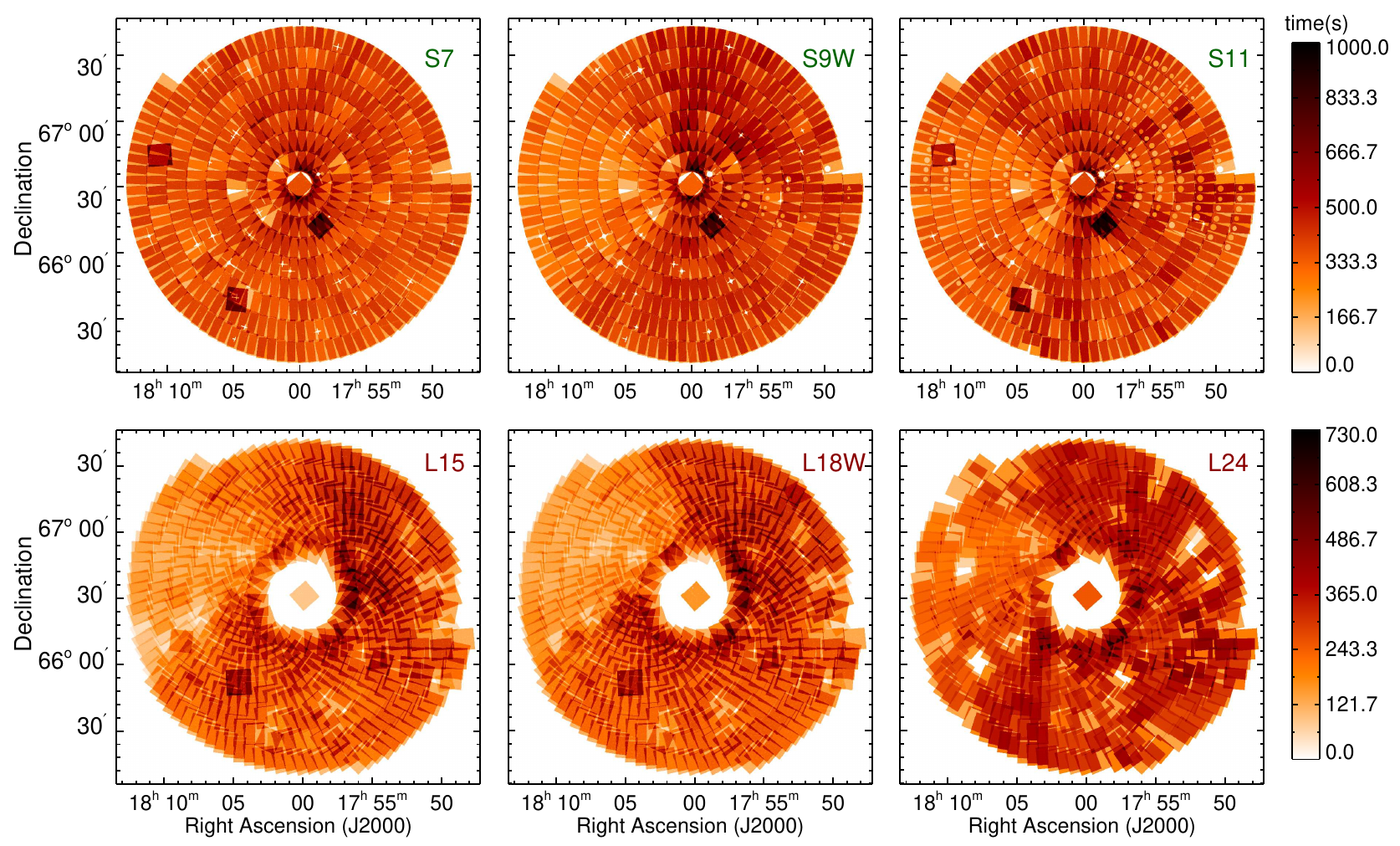}}
\caption{Entire coverage maps of MIR-S (upper) and MIR-L (lower) bands.}
\label{fig 10}
\end{center}
\end{figure*}

 To decide the area to be masked, we had to trace the bright sources causing MUXbleeds. We 
investigated the threshold brightness at which MUXbleeds started to be created. We carried out 
photometry on all of the individual frames to find the sources brighter than this threshold. The 
photometry was done both before and after the L.A cosmic procedure  to search for the sources hit by 
cosmic rays and deformed during the cosmic ray rejection process. The results of these measurements 
enabled us to determine the area to be masked, as well as the sources distorted by cosmic rays. 
The typical shape of the masked region around a bright source is a circular disk centered on the 
centroid of the source, with a narrower horizontal stripe covering the bleeding trail. We found 
that sources brighter than $\sim$ 12.6 magnitude (mag) cause MUXbleeds in the $N2$ band, and 
sources brighter than about 13 mag in the $N3$ and $N4$ bands. The radius of the circular area and 
the width of the stripe both depend on the brightness of the source. For the circular mask, the 
center is defined by the pixel coordinate of the source that causes the bleeds, and the appropriate 
radius was chosen to be 2.5 times of the Kron radius (Kron 1980). The width of the stripe was set 
to be 1.5 Kron radius, and the actual position of stripe was 2 -- 3 pixels parallel-shifted in a 
vertical direction from the original y-coordinate of the source to efficiently block the MUXbleed, 
because it is not symmetric with respect to horizontal axis. Applying these criteria on each frame, 
we masked the MUXbleeds and extremely bright sources (brighter than 12.6 mag) as well as the 
sources deformed by the cosmic ray rejection procedure in the NIR bands.

In the MIR bands, the MUXbleeds have less significant effects on the image frames because the 
bleeding trails are rarely found in the $S7$ and $S9W$ bands, and completely disappear in the $S11$ 
band. Since the horizontal trails usually do not extend to the edge of frame, we used small patches 
covering the trails around the source in order not to lose too many pixels. However, in 19 frames 
in the $S9W$ and 104 frames in the $S11$ bands, a bright bean-shaped pattern remained in the corner 
of the frames. Artifacts occasionally caused by moving objects such as satellites or asteroids 
passing through the field of view were also found in four frames in the MIR-S bands. To reject 
these bright artifacts, masking regions were carefully determined by visually inspecting the 
individual frames and  weight maps were generated that were used for image mosaicking basically 
in the same manner as the NIR data. The number of frames that requires a weight map for mosaicking 
was about 10$\%$ of the MIR-S data.

\subsubsection {Weight maps and mosaic images} 

 To create a mosaic image for each band, we combined individual frames using SWarp. During the 
SWarp run, the `WEIGHTED' option was selected as a combine type in order to use the weight maps 
generated for MUXbleed correction.  The BILINEAR resampling method was also used since it is 
known to be effective at suppressing boundary discontinuities. In the final mosaic images, spiky 
structures still remained close to bright sources because they are  difficult to remove unless 
we apply rather large circular radii when masking these regions. However, these structures do 
not occupy a significant fraction and do not cause serious false detection problems. False 
sources are easily filtered out during the confirmation procedure that uses data from other 
bands (see \S4).

\begin{figure*}[ht!]
\begin{center}
\resizebox{\hsize}{!}{\includegraphics{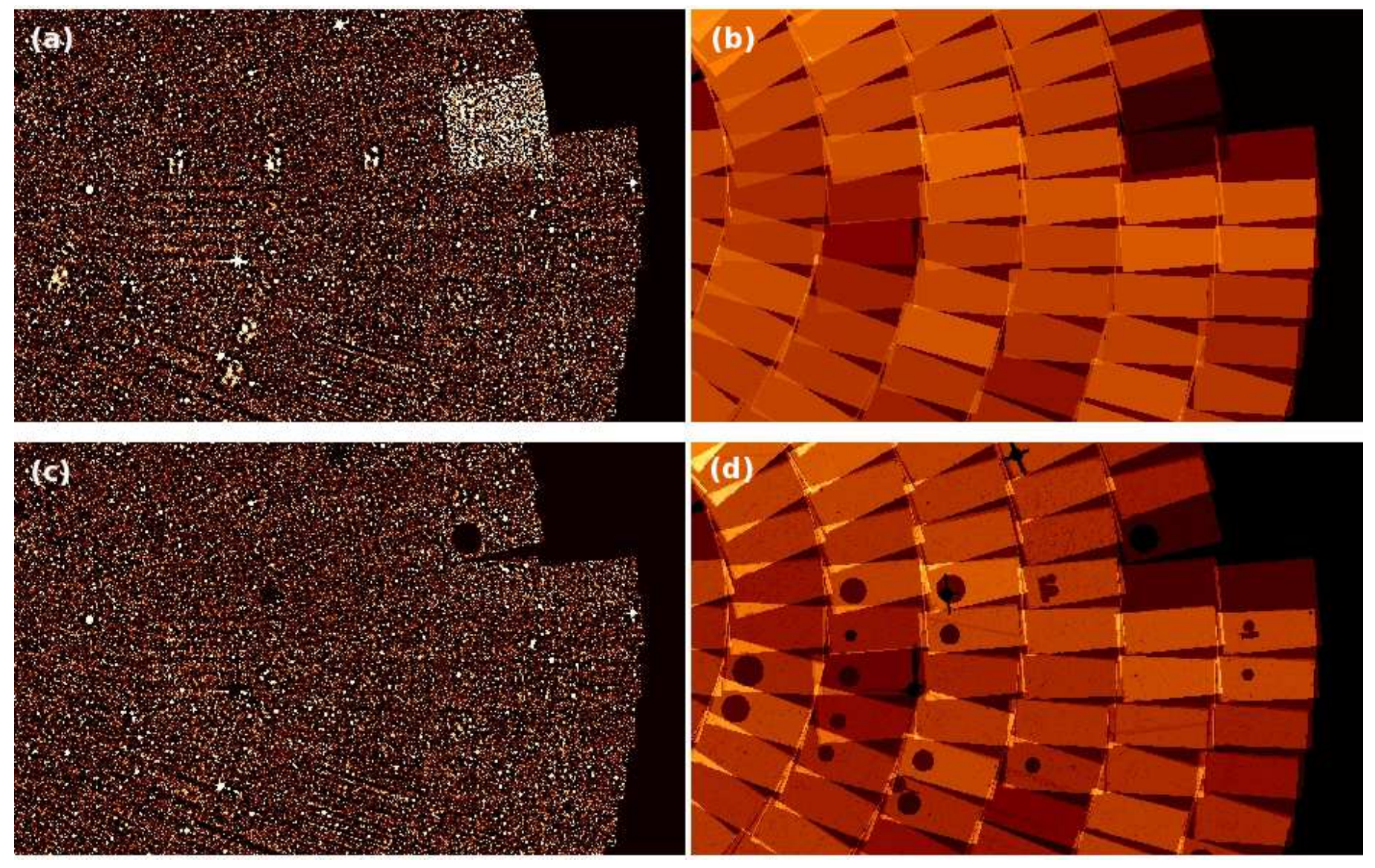}}
\caption{Segments of a mosaic images (left) and weighted coverage maps (right) before and after
the correction for undesirable patterns in the $S9W$ image. The upper panels show the images
before the masking and the bottom panels show the resultant images after masking.  Coverage maps
show that the masking of MIR-S images effectively removes most of the remaining patterns (for
19 frames), and demonstrates that remaining MUXbleed effects are almost completely removed.}
\label{fig 11}
\end{center}
\end{figure*}

\begin{table*}[]
\tabletypesize{\small}
\centering
\caption{. Coverage and masked area }
\label{table:1.}
\begin{tabular}{ccccccccccccc}
\hline\hline
\multirow{3}{2cm}{IRC bands } & \multicolumn{3}{c}{NIR} & \ & \multicolumn{3}{c}{MIR-S} & & \multicolumn{3}{c}{MIR-L}  &   \\
      \cline{2-4}  \cline{6-8} \cline{10-12}\\
    &  {N2} & {N3}  &{N4} & & {S7} & {S9W} & {S11} & & {L15} & {L18W} & {L24} \\
\hline
 Number of combined frames     &  443  &   443  &  443  & &  445  &  445  &  445  & & 445  & 445   & 403  \\
 Area (deg$^2$)                & 5.338 & 5.338  & 5.335 & & 5.330 & 5.341 & 5.352 & & 4.983& 5.007 & 4.941\\
 Masked area (deg$^2$)         & 0.214 &  0.198 & 0.111 & & 0.002 & 0.026 & 0.016 & & 0.0  & 0.0   & 0.0  \\
 Fraction of masked area ($\%$)&  4.0  & 3.7    &  2.1  & &  0.04 & 0.5   & 0.3   & & 0.0  & 0.0   & 0.0  \\
\hline
\end{tabular}
\end{table*}

Fig. 8 shows the results of the image correction for MUXbleed effects using two pairs of maps.
The upper panels are mosaic (left) and weight (right) maps before correction, whereas the bottom 
panels show the corresponding images after correction. Most of the bleeding trails in the NIR 
bands are effectively removed by this process. By co-adding the individual pointed observations,
utilizing the weight map for each frame, we finally produced three NIR master images for the source 
extraction and photometry. Among 446 frames, three were excluded owing to the stacking failure in
the pipeline caused by the poor data quality. The actual area covered by the NIR bands is about 
5.34 deg$^{2}$ and the fraction of masked region is in the range of 2 -- 4$\%$ of the observed 
area. As summarized in Table 1, about 0.21 deg$^{2}$ was masked out in the $N2$ band, which is 
about 4$\%$ of the entire area covered by the $N2$ band. In the $N3$ band, 0.198 deg$^{2}$ (3.7$\%$)  
and in the $N4$ band, 0.111 deg$^{2}$ (2.1$\%$) were masked, respectively.

The weighted coverage maps may help us to understand  the overall view of the final 
results compared to the uncorrected image data, as shown in Fig. 9. The maps for the NIR data
before and after the correction  are presented for comparison. The actual coverage as well as the
masked area were calculated using these maps.  The mosaic images for the MIR bands were generated 
in the same way using weight maps for individual frames, and have similar coverage maps as shown 
in Fig. 10.

The areas observed by the MIR-S bands range between 5.33 and 5.35 deg$ ^2$.  The fraction of the 
masked region (zero-weighted regions) is less than $\sim$ 0.03 $\%$, a very small fraction compared 
to those of the NIR bands (Table 1). The masking of the $S7$ band was mainly required because of the 
streaks caused by bright stars in $\sim$ 25 frames. For the $S9W$ and $S11$ bands, the masking was 
done to reject the noticeable patterns in about 19 and 104 frames in the $S9W$ and $S11$ bands.  
Fig. 11 shows the segments of resultant mosaic images of the $S9W$ band.  The upper and lower panels 
show the images before and after the correction for the irregular artifacts, respectively.  For the 
MIR-L bands, this procedure was not necessary because there were no artifacts. Note that there is 
uncovered region near the central parts of the MIR-L observation because of the offset between the 
FoVs for the MIR-S and MIR-L observations (see Murakami et al.  2007, for the focal plane allocation 
of the instruments of AKARI).

\section {PHOTOMETRY  AND DATA PROPERTIES}

\subsection {Source detection and photometry}

As shown in Fig. 8 and Fig. 11, we removed most of the MUXbleed effects in the NIR bands, as well 
as other artifacts in the MIR bands, to minimize spurious detections. We confirmed that the detection 
reliability was significantly improved by comparing the detected sources from a small portion of both 
the corrected and the uncorrected images, using the same detection parameters. 

We used the entire mosaicked images covering the whole NEP-Wide area to carry out extraction and 
photometry of the sources for each band. To measure the fluxes of the detected sources, we used a 
software SExtractor developed by Bertin \& Arnouts (1996) \footnote{For the detailed description of 
this software, See http://terapix.iap.fr/IMG/pdf/sextractor.pdf}. Here, we chose DETECT\_THRESH$=$3, 
DETECT\_MINAREA$=$5, and BACK\_SIZE=3,  which are the same as those used by Wada et al. (2008), 
except for DETECT\_THRESH. We  chose a higher threshold to reduce  false detections.  The number 
of detected sources from nine master images are presented in Table 2, together with the estimated 
detection limits as described in the following section. About 87,800, 104,000, and 96,000 sources 
are detected in the $N2$, $N3$, and $N4$ band, respectively. The numbers of detected sources in 
the MIR bands are much smaller than those in the NIR bands. In the MIR-S bands, 15,300 ($S7$), 
18,700 ($S9W$), and 15,600 ($S11$) sources were detected, and in the MIR-L bands, 13,100 ($L15$), 
15,100, ($L18W$), and about 4,000 ($L24$) sources were detected, respectively.

The photometric measurements were made in a single mode operation for each band in order not to use 
the same aperture for different band images. The sizes of the sources depend  significantly on the 
effective wavelengths of the filter bands  because the IRC images are nearly diffraction limited. 
To confirm the validity of detected sources and  reject spurious objects, it is more appropriate 
to employ the single mode operation for each band and look for counterparts in the other bands. 

 To use flexible apertures for various sources, the fluxes of them were measured using elliptical 
Kron apertures (i.e., SExtractor's Flux\_AUTO), owing to the elongated shapes of the PSFs in the NIR 
bands as well as for the variable sizes and shapes of MIR band sources.  The fluxes in units of  
ADU are converted to $\mu$Jy using the flux calibration in Table 4.6.7 of the IRC data user manual 
version 1.4 (Lorente et al. 2008; Tanabe et al. 2008), which has been established based on the 
observation of  standard stars. Finally, we obtained the AB magnitude (Oke \& Gunn 1983) using 
the relation, AB (mag) = $-2.5$ log $f_{\nu} + 23.9 $, where $f_{\nu}$ is the flux density within 
a given passband in units of $\mu$Jy.

\begin{deluxetable}{cc ccc c ccc c ccc}
 \tabletypesize{\small}
 \rotate
 \tablecaption{\label{tab:phot} Number of detected sources and 5$\sigma$ detection limits}
 \tablewidth{0pt}
 \tablehead{
 \multirow{3}{2cm}{IRC bands }
   & \multicolumn{3}{c}{NIR} & \colhead{} &
  \multicolumn{3}{c}{MIR-S} & \colhead{} & \multicolumn{3}{c}{MIR-L}  & \colhead{}  \\
 \cline{2-4}  \cline{6-8} \cline{10-12}\\
  & \colhead{N2} & \colhead{N3}  &\colhead{N4} & & \colhead{S7} & \colhead{S9W} &
  \colhead{S11} & & \colhead{L15} & \colhead{L18W} & \colhead{L24}
  }
\startdata
FWHM of PSF ($^{''}$) & 4.8 &4.9  & 4.9 & &5.8 & 5.9& 6.1 & & 6.5 & 6.9 & 7.3\\
Number of detected sources & 87,858 & 104,170 & 96,159 & & 15,390 & 18,772 & 15,680 & & 13,148 & 15,154 & 4,019 \\
Detection limit in AB      &  20.93 & 21.09 & 21.07 & & 19.48 & 19.33 & 18.97 & & 18.59 & 18.70 & 17.82  \\
  (in $\mu$Jy)             & (15.42)&(13.30)&(13.55)& &(58.61)&(67.30)&(93.76)& &(133.1)&(120.2)&(274.4) \\
50\% completeness in AB    & 19.75  & 19.81  & 19.87  & & 18.7  & 18.6  & 18.2  & & 17.9  & 18.0  &  16.8 \\
   (in $\mu$Jy)            &(45.68) &(43.39)&(41.02)& &(120.2)&(131.8)&(190.5)& &(251.2)&(229.1)& (691.8)\\
\enddata
\end{deluxetable}

\begin{figure*}[ht!]
\centering
\subfigure{\resizebox{.95\hsize}{!}{\includegraphics{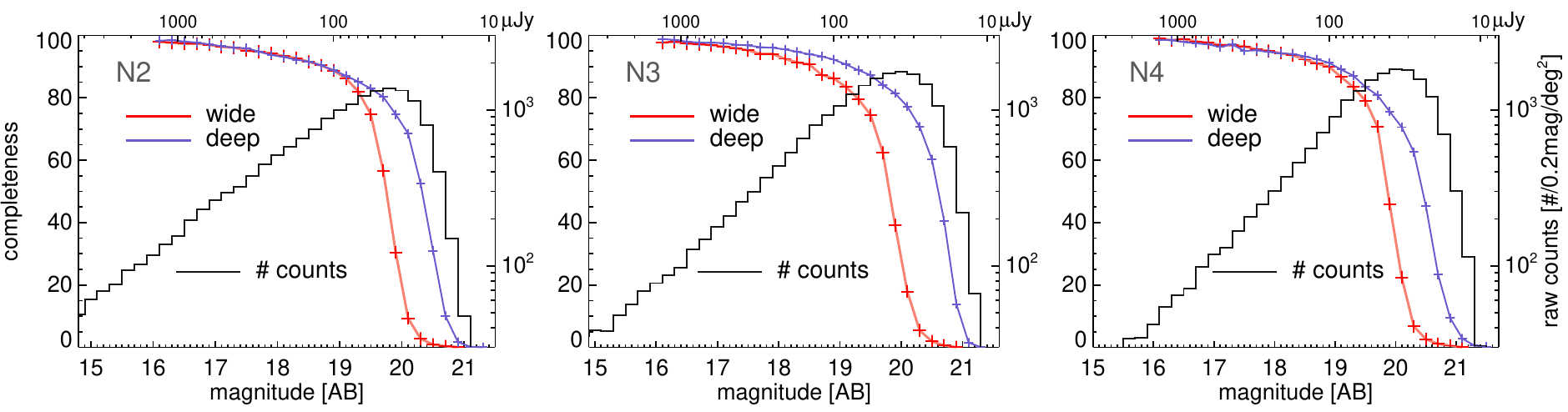}}}
\subfigure{\resizebox{.95\hsize}{!}{\includegraphics{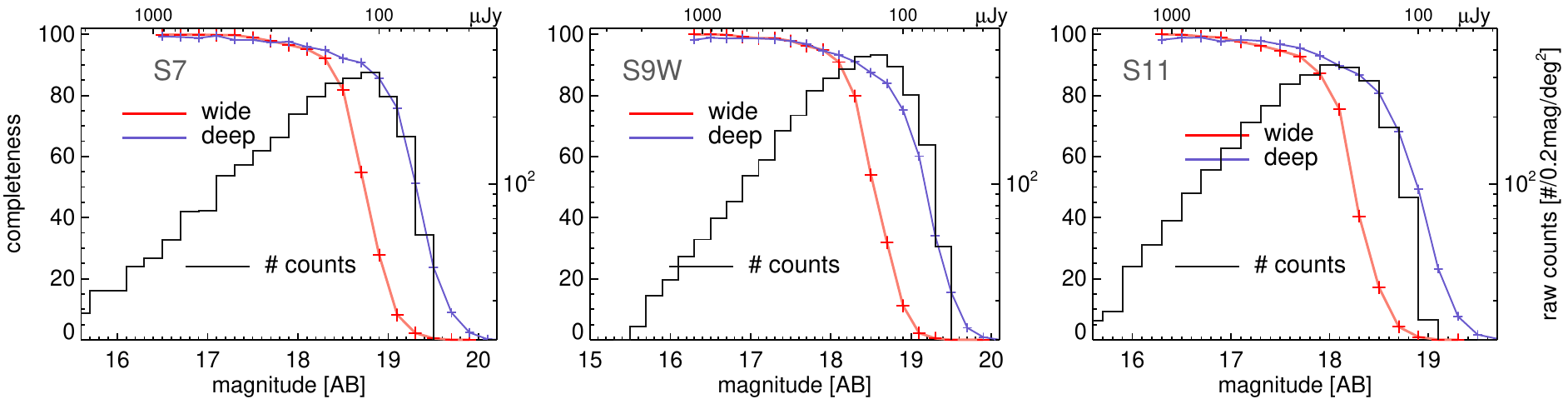}}}
\subfigure{\resizebox{.95\hsize}{!}{\includegraphics{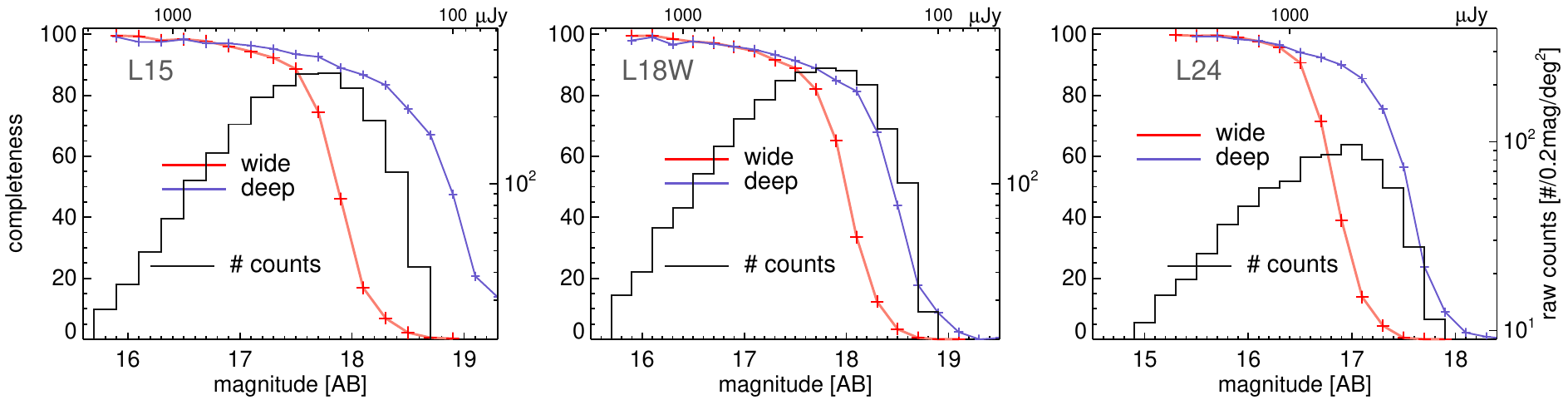}}}
\caption{The completeness estimation for each band (red curves). The estimates for the NEP-Deep data
are also shown for comparison. The gray histograms show the NEP-Wide source density per square degree
per 0.2 magnitude bin at each band (not corrected for incompleteness). }
\label{fig 12}
\end{figure*}

We checked the reliability of the photometry by comparing our magnitudes of the  bright ($<$ 16 
mag) sources with those in the NEP-Deep catalog of Wada et al. (2008, see also Takagi et al. 2012). 
We found that the average magnitudes of the same sources in the NEP-Wide and  NEP-Deep differ by 
up to 0.05 mag. Since the rms of magnitude differences between the NEP-Deep and Wide data were about 
0.1, the systematic  difference of 0.05 magnitude is not considered to be statistically significant.
Furthermore, these differences are within the absolute calibration uncertainties ($\sim6\%$). 
Considering the observations for the NEP-Wide and NEP-Deep surveys were done with different observing 
templates (i.e. IRC03 for NEP-Wide and IRC05 for NEP-Deep) and the photometry was performed with 
slightly different parameters, we regard that the small systematic differences between the measured 
magnitudes are not serious.

\subsection {Detection limits and completeness}

The flux limit of the point source detection in each band was estimated from the fluctuation in the 
sky background, by measuring the flux at random positions far away from the source positions. We used 
an aperture three times the size of the FWHM, and determined the 5$\sigma$ detection limits based on 
the value of $\sigma$ derived from the sky background.  The detection limits depend on the noise levels 
of the fields, which vary from place to place, and we present the averaged values over the entire NEP 
field in Table 2.  In this table, we listed  the FWHMs of sources detected in mosaicked images for all 
the IRC bands.  For each band, we measured the FWHM for about 30 bright sources whose optical 
counterparts have stellarities greater than 0.95 and took the average of them.  In the NIR bands, the 
$N2$ filter reaches a depth of $\sim$ 20.9 mag, and  the $N3$ and $N4$ bands  reach $\sim$ 21.1 mag. 
The MIR detection limits are much shallower:  $\sim$ 19.5 ($S7$), 19.3 ($S9W$), and 18.9 mag ($S11$) 
for the MIR-S bands, and $\sim$ 18.5 ($L15$), 18.6 ($L18W$), and 17.8 mag ($L24$) for the MIR-L bands.
In the table, the $50\%$ completeness limits, which were  measured by injecting artificial sources as 
described below, are also presented.

\begin{figure}[ht!]
\begin{center}
\resizebox{\hsize}{!}{\includegraphics{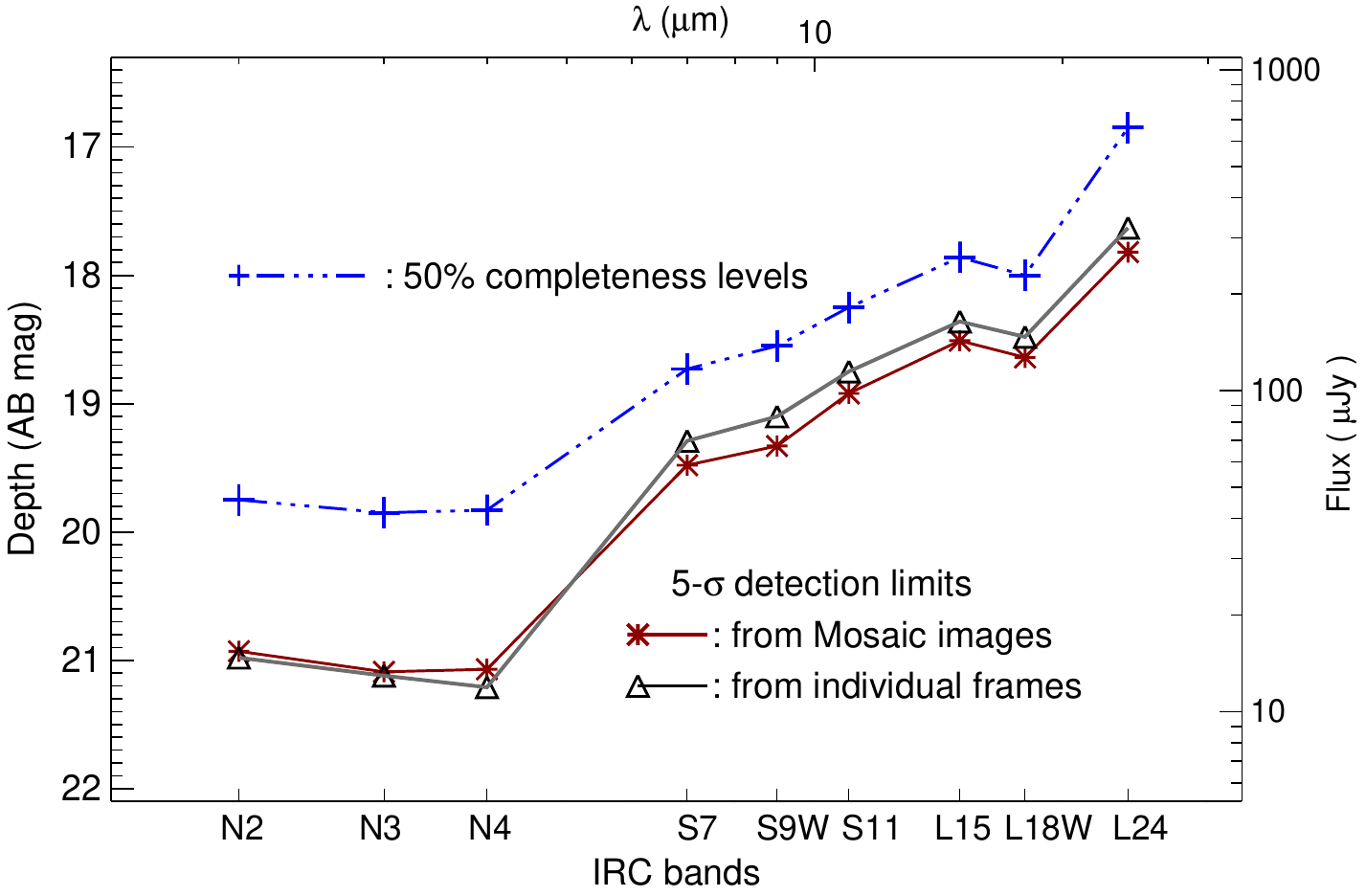}}
\caption{The flux detection limits and 50$\%$ completeness limits for each band. The detection limits
of the NIR bands are around 21 mag, and MIR bands reach much shallower depth than NIR bands. The
50$\%$ completeness limits are about 1.2 mag shallower than the detection limits in the NIR filters
and about 0.7 mag shallower in the MIR bands from $S7$ to $L18W$.}
 \label{fig 13}
\end{center}
\end{figure}

Using the IRAF tasks in {\texttt{noao.artdata} package, we generated artificial sources with a fixed 
range of magnitude and spread them at random positions in 8 sample regions of 10$^{\prime}\times 
10^{\prime}$ selected from all over the mapped area.  The sources injected at positions within a 
distance of 20 pixels from any other sources are not counted as input sources  to avoid source 
blending and miscounting. We attempted to detect the injected objects and measured the brightness 
of them by running SExtractor with the same parameters as those applied to the detection and 
photometry for the real sources.  We compared the positions and magnitudes of the detected sources 
with those of the artificial input sources. If a detected source lay within 2 pixels (1 pixel 
corresponds to 1.46$^{''}$, 2.34$^{''}$ and 2.45$^{''}$ for NIR, MIR-S, and MIR-L bands, respectively) 
from the location of the input source, and the measured brightness was within 0.5 mag of the input 
value, we regarded this source as valid detection of the injected source. These criteria for the 
validation of the detected sources are the same as those chosen by Wada et al. (2008).  We then 
counted the number of detected sources and compared them with the number of the input artificial 
sources to determine the completeness, which was defined as the fraction of recovered objects in each 
magnitude bin.  To ensure statistical efficacy, we generated a sufficient number of input sources (300
-- 400) per sample region.  For each magnitude bin, we repeated the same procedures seven to ten 
times with different seed numbers for the brightness and spatial distributions, and took the average. 
Among the various input parameters for this test, the most sensitive was found to be the half 
intensity radius at each wave band.

The completeness of detection for the NEP-Wide data are shown in Fig. 12 as a function of magnitude 
for the IRC bands. We also show the completeness estimates for the NEP-Deep data using the same 
parameters as those employed for the NEP-Wide data.  The completeness curve shows that the detection 
probability begins to drop rapidly at around the 85 -- 80$\%$ value in the NIR bands, and around the 
90$\%$ level in the MIR bands. The magnitude difference between the 90$\%$ and 10$\%$ completeness is 
about 1 mag in the NIR bands, and less than 1 mag in the MIR bands. The 50$\%$ completeness limits are 
about 19.8 mag in the NIR bands, 18.7 -- 18.3 mag in the MIR-S bands, and 18.0 -- 16.8 mag in the MIR-L 
bands, as presented in Table 2. As shown in Fig. 12, the magnitude differences between the NEP-Wide 
and NEP-Deep data at the 50$\%$ completeness level are about 0.5 -- 0.6 mag in the NIR and MIR-S bands. 
In the MIR-L bands, the differences are about 1.0 ($L15$) -- 0.5 mag ($L18W$).

Fig. 13 shows the comparison of the 5$\sigma$ detection and $50\%$ completeness limits for the IRC 
bands. The measurements of the detection limits using individual frames are also shown for comparison 
(gray). These estimates obtained using the mosaicked images give slightly shallower detection limits 
than those measured using the individual frames for the NIR band, while deeper limits are measured 
for the MIR bands. The differences at the NIR bands are less than 0.2 mag and possibly result from 
the variation of the image quality, for example from FWHM changes and seeing variations, along with 
resampling during the mosaicking process.

The 50\% completeness limits are shallower than the 5$\sigma$ detection limits by about 1.2 -- 1.3 mag. 
for the NIR bands and about 0.6 -- 0.8 mag for the MIR bands (except for $L24$ which has a relatively 
larger difference of 1.0 mag.) Wada et al. (2007) also found a similar trend in the differences 
between the 5$\sigma$ detection limits and 50\% completeness limits for the NEP-Deep data. By comparing 
a single exposure and stacked image of ten exposures, they found that the 50\%  limits do not 
improve much for the NIR bands, while the improvements for the MIR bands are  close to the square-root 
of the exposure time.  On the basis of these results, they concluded that the NIR bands are affected 
by the source confusion.  In fact, the improvement in the 50\% limit between one pointing and ten 
pointing observations is only  a factor of 1.15 (Wada et al. 2007) for the $N3$ band, implying that
the source confusion could be significant. 

The MIR bands are unlikely to be affected by the confusion because the source density is much lower 
than that of the NIR bands, while the sizes of the PSFs are nearly the same (Lee et al. 2009). The 
smaller difference between the 50\% completeness and 5$\sigma$ detection limit can thus be
understood by the confusion effects in the NIR bands.  Note that the numbers of source per beam are 
1/60.7, 1/46.8, and  1/51.3  for $N2$, $N3$ and $N4$ bands, respectively.  These values are smaller 
than the classical definition of the confusion limit of 1/30 sources per beam, but only by a factor 
of two.  The NIR band observations are affected by the source confusion to some extent.

\section{CONFIRMATION OF THE SOURCES}

\subsection{Supplementary data}

In addition to our AKARI/NEP-Wide data, high-quality optical data, NIR $J$ and  $H$ band data, and 
radio data over a more limited field are available for the NEP-Wide field (Kollgaard et al.  1994, 
Lacy et al. 1995, Sedgwick et al. 2009, White et al. 2010). Optical data were 
obtained using the 3.5m Canada-France-Hawaii Telescope (CFHT) for the inner parts and the 1.5m 
telescope at Maidanak observatory in Uzbekistan for the outer parts of the NEP-Wide field as shown 
Fig. 1.  The CFHT observations with the MegaCam covered the inner part of the 2 deg$^2$ rectangular 
field centered on the NEP using the $u^{*}$, $g^{'}$, $r^{'}$, $i^{'}$, $z^{'}$ filter system. The 
detection limits (4$\sigma$) are about 26 mag for $u^{*}$, $g^{'}$, $r^{'}$, about 25 mag for $i^{'}$, 
and 24 mag for $z^{'}$ band, and the full catalog contains over $\sim$ 110,000 sources (Hwang et al. 
2007). The Maidanak observations were carried out using the SNUCAM (Im et al. 2010). The observations 
covered the outer regions surrounding the CFHT field using $B$, $R$, $I$ filters (Jeon et al. 2010),
whose depths are around 23 mag in B, R and about 22 mag in I band.  In addition, NIR $J$, $H$ band 
data was obtained using FLAMINGOS mounted the Kitt Peak National Observatory (KPNO) 2.1m telescope, 
which cover the entire NEP-Wide area ($\sim$ 5.2 deg$^{2}$). The number of sources in this data is 
over 220,000 (Jeon et al., in preparation). The optical data are crucial for identifying the nature 
of the corresponding AKARI sources, since the stellarity parameters help us to distinguish between 
stars and galaxies. The $J$, $H$  data are used to bridge the gap in wavelength coverage between 
the AKARI/NIR and the optical data.

\subsection{Overview of the source matching}

To construct a reliable source catalog, we have to validate the detected sources. If a certain source
is detected at only one filter band without any counterpart in the other bands, it could potentially
be as a false detection, especially in the case of the NIR bands. In order to verify the reliability
of detected sources in a given IRC band, we searched for their counterparts within a 3$^{''}$ radius
in the other IRC bands, as well as  ancillary optical and $J$, $H$ band data search parameter.
The choice of 3$^{''}$ as a matching radius is somewhat arbitrary: we considered the astrometric
accuracy of the NEP-Wide data to be  1.38$^{''}$ (Lee et al. 2009), hence selected a search radius of
twice of this value.  We tried several values and found that the number of matched sources begins to
increase very slowly with radius larger than 2$^{''}$, and nearly saturates at around 3$^{''}$.
Note that the typical size of the FWHM of the point sources ranges from 5$^{''}$ (NIR) to 7$^{''}$ (MIR)
(see Table 2), approximately corresponding to the diameter of the matching circle.

\begin{figure}[ht]
\centering
\resizebox{1.05\hsize}{!}{\includegraphics{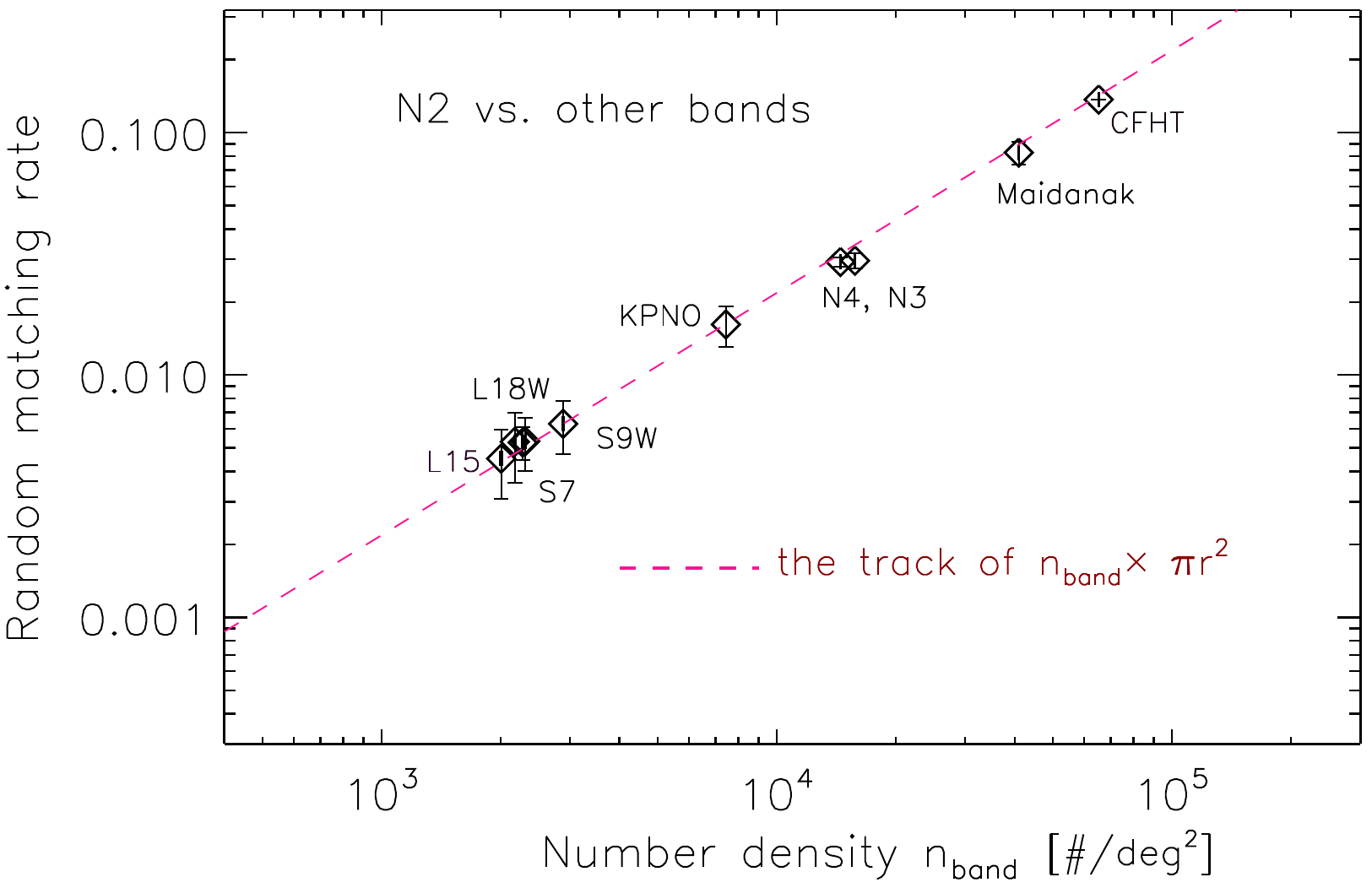}}
\caption{
The probability of random matching for the $N2$ band sources with those in other bands. The random
matching probability is proportional to the density of the sources in the band to be matched, as
indicated by the dotted line.
}
\label{fig 14}
\end{figure}

\begin{deluxetable}{cc ccccccccccc}
 \tabletypesize{\tiny}
 \rotate
 \tablecaption{\label{tab:phot} The number of sources matched with those in other bands}
 \tablewidth{0pt}
 \tablehead{ \colhead{IRC Bands }
     & \colhead{Optical}  & \colhead{N2} & \colhead{N3}  &\colhead{N4} & \colhead{S7} & \colhead{S9W} & \colhead{S11}
        &\colhead{L15} & \colhead{L18W} & \colhead{L24} }
 \startdata
  N2  & 72,871 (83$\%$) & 87,858  (0) &  81,012 (22) & 71,919 (21) &  13,357 (1) & 16,017 (3) & 12,643 (2) &  8,170 (1) &  8,554 (2) & 2,070 (1)\\
  N3  & 77,223 (74$\%$) & 81,049 (59) & 104,170 (0)  & 84,856 (49) &  13,752 (7) & 16,563 (5) & 13,180 (3) &  9,321 (2) & 10,167 (3) & 2,266 (0)\\
  N4  & 67,667 (70$\%$) & 71,938 (40) &  84,850 (43) & 96,159 (0)  &  14,169 (7) & 16,996 (7) & 13,434 (2) &  9,784 (5) & 10,841 (6) & 2,454 (0)\\
  S7  & 13,928 (90$\%$) & 13,356 &  13,745 & 14,162 &  15,390 & 12,091 &  8,489 &  4,790 &  4,750 & 2,059  \\
  S9W & 16,662 (89$\%$) & 16,014 &  16,558 & 16,989 &  12,091 & 18,772 & 12,923 &  6,757 &  7,040 & 2,389  \\
  S11 & 13,404 (85$\%$) & 12,641 &  13,177 & 13,432 &   8,489 & 12,923 & 15,680 &  7,084 &  7,229 & 2,476  \\
  L15 &  8,628 (66$\%$) &  8,169 &   9,319 &  9,779 &   4,790 &  6,757 &  7,084 & 13,148 &  9,377 & 2,571  \\
 L18W &  9,217 (60$\%$) &  8,552 &  10,164 & 10,835 &   4,750 &  7,040 &  7,229 &  9,377 & 15,154 & 2,673  \\
  L24 &  2,280 (57$\%$) &  2,069 &   2,266 &  2,454 &   2,059 &  2,389 &  2,476 &  2,571 &  2,673 & 4,019 \\
\enddata
\end{deluxetable}

In the case of the $N2$ band, for example, we first examined its positional matching with the $N3$ and
$N4$ band. We then proceeded to find counterparts in the optical, and the KPNO's $J$ and $H$ bands.
We finally looked for the matching sources in the AKARI's MIR-S and MIR-L data. About 1,590 of the $N2$
sources ($\sim$ 1.8$\%$) do not have any counterpart in any of the other bands. In the $N3$ and $N4$
bands, about 3.9$\%$ and 6.8$\%$ of the sources remained unmatched to any other band data,
respectively. All of these unmatched sources can not be confirmed and are likely to be false objects.
We excluded these sources from the catalog.

To find the counterparts of the MIR sources, a similar procedure was applied. The cross-matching of the 
sources in each of the MIR band was carried out against the others from the NIR to MIR-L bands in order 
of wavelength. The fractions of sources without any counterparts are 1.6$\%$, 1.0$\%$, and 2.2$\%$ for 
the $S7$, $S9W$, and $S11$, respectively. For the MIR-L bands, about 7$\%$, 10$\%$, and 20$\%$ of 
sources in the $L15$, $L18W$, and $L24$ bands remained unmatched to  those in other bands. Unlike 
the NIR bands, it is risky to exclude all of the MIR sources having no counterparts because the 
artifacts are not serious in the MIR bands compared to the NIR bands, and very red objects can be 
detected in the long wavelength bands. We therefore checked all the unmatched sources by eye and 
included most of them except for some very rare cases ($\sim 2\% $) with spurious images.

The results of the matching procedures among the IRC bands as well as  with  optical data are 
presented in Table 3. The numbers in this table include the sources that were matched more than once, 
which are given in the parenthesis. Duplicate sources, or multiple matching occurred predominantly 
when the sources were located in a crowded region, although numerically they represent a small fraction 
of the total. When two band data are used to find a counterpart, there is always a possibility of false 
matching. Suppose that the sources in bands $A$ and $B$ are uncorrelated and randomly distributed in 
space. In this case, the fraction of the sources in band $A$ accidentally matched with those in band 
$B$ is simply $n_B \times \pi r_{match}^2$, where $r_{match}$ is a matching radius.  We confirmed this 
simple relationship between the source density and the false matching rate by numerical simulations: 
we tried to match the $N2$ sources to those in other band data but in different fields of the same size, 
the results of which are shown with diamonds in Figure 14.  A dotted line in this figure is an 
expected relation from the numerical simulation. The matching experiments with the real data for random 
fields of view show good agreements with the expected relation based on a random-matching probability 
argument.  In our cross identification procedure, the density of the optical data is the highest: 
the CFHT catalog (Hwang et al. 2007) contains about 118,200 sources in 1.8  deg$^2$ and thus the 
source density is 65,700 per deg$^2$. When we used the matching radius of 3$^{''}$, the false 
matching rate was about 14.3\%. For the Maidanak data, the source density is about 41,100 per deg$^2$ 
and the false matching rate is estimated to be 9\%.

However, these probabilities should be regarded as upper limits for the false matching since the
sources detected in the same field of view  should be actually correlated, i.e., a source at a given
band is likely to be present in the other bands.  In order to make more realistic estimates of the 
false matching rate, we varied the matching radius.  In the case of the matching test between the $N2$ 
and CFHT optical data, for example, we found that the number of the $N2$ sources that have the optical 
counterparts in an annulus corresponding to matching radii of between 2.5$^{''}$ and 3$^{''}$ comprises 
about 1.41\%  instead of the random matching probability of 4.37\%.  In the case of the matching with
the Maidanak data, we expect a 2.75\% false matching rate for the same annulus, but only 1.08\% 
matched sources were found.  If we assumed that all the sources in this annulus had been accidently 
matched, the false matching could be less than 1.5\%. We thus conclude that the false matching 
probability must be very low.

\subsection{Summary of the source matching}

\subsubsection{Matching between IRC bands}

\begin{figure}[ht!]
\begin{center}
\resizebox{\hsize}{!}{\includegraphics{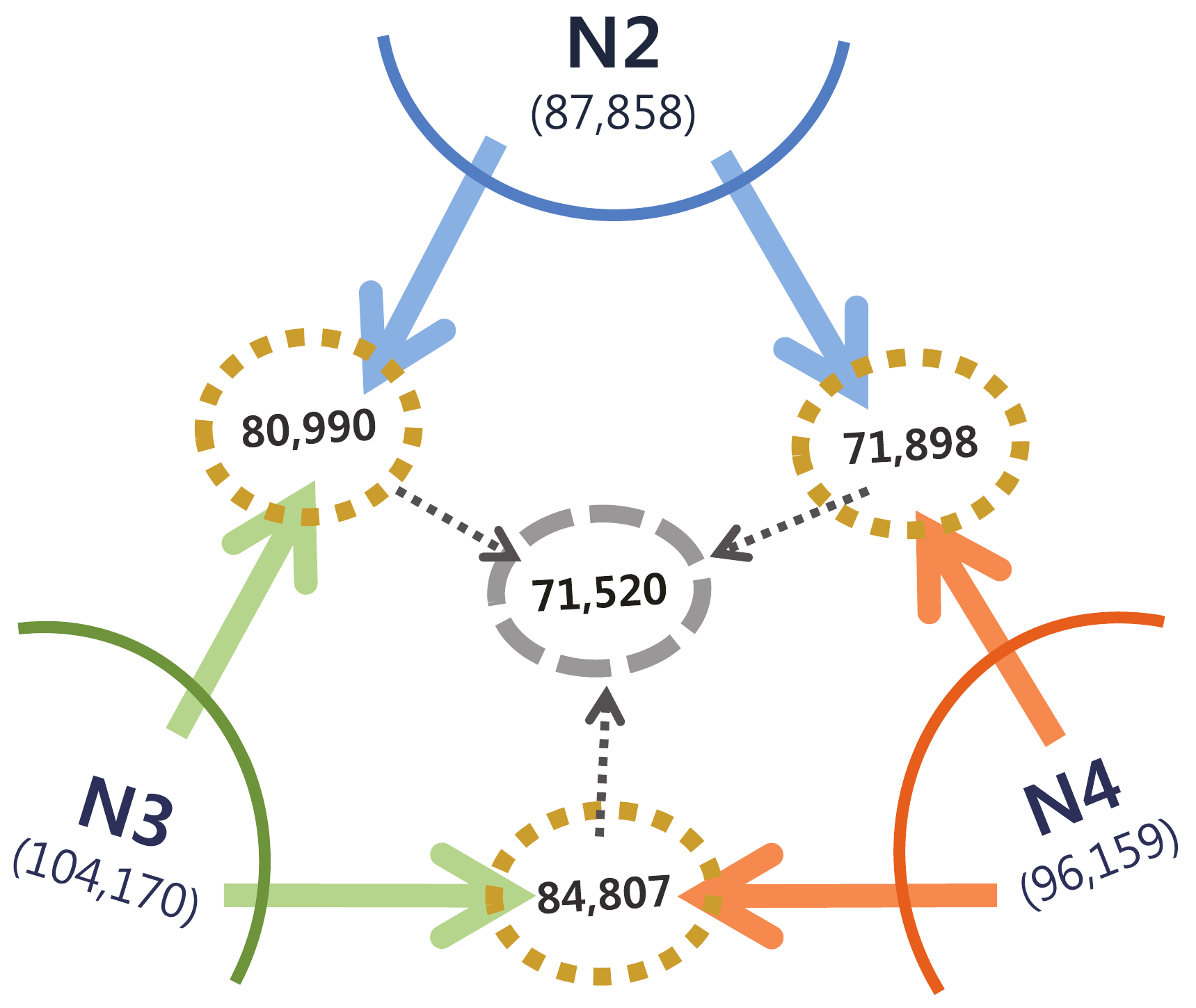}}
\caption{Summary of the source matching between the different NIR bands. The numbers of sources
detected in each NIR band are presented at the apexes of the triangle. The matching results amongst
them are shown, excluding any duplicated or multiply matched sources. The numbers in the dotted
ellipses represent the number of sources with  detections  in both of the matching bands.}
\label{fig 15}
\end{center}
\end{figure}

The results of the source matching amongst the different wavebands were used to confirm the source 
detections. In the AKARI/NEP-Wide data, the numbers of sources in the NIR bands are much larger than 
those of the MIR band sources. Consequently, the majority of the MIR sources have the NIR counterparts, 
but not vice versa.  It is therefore reasonable to combine the information on the basis of the NIR 
framework.

\begin{figure*}[ht!]
\centering
\resizebox{0.9\textwidth}{!}{\includegraphics{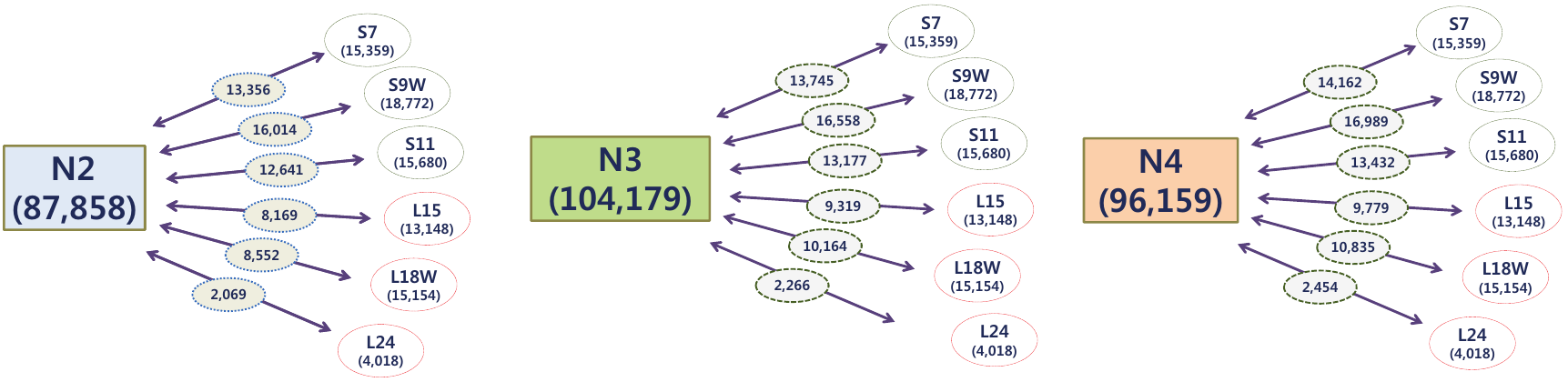}}
\caption{Schematic diagram showing the results of the matching between the sources in the NIR and MIR
bands except for multiply matched sources. The matching results of the $N2$ sources with all the MIR
bands are shown in the leftmost panel. The results for $N3$ and $N4$ bands are shown in the middle
and the right panels, respectively.}
\label{fig 16}
\end{figure*}

As explained in the previous section, a matching between any two bands was performed in both 
directions. Between the $N2$ and $N3$ bands, for example, we first searched for  counterparts in the 
$N3$ band for a given $N2$ source, and then changed the matching direction. Similar procedures were 
applied to all the possible combinations of the three NIR bands, as shown in Fig. 15. The numbers 
shown in the ellipses on the three sides of the triangle are those of the number of sources, with the 
detections in both bands indicated at the apexes. Using the results of the cross-matching in both 
directions, we can find either duplicate or multiply matched sources between two bands. The number 
of the sources detected in both the $N2$ and $N3$ was 80,990, in  the $N2$ and $N4$ was 71,898, and 
in the $N3$ and $N4$ was 84,807.

For a given MIR source, there could be multiple NIR or optical sources since the PSF of the MIR data 
is much larger than that in the NIR and the source density in the NIR is much higher (by up to 
a factor of eight) than that in the MIR. In such cases, we simply chose the nearest detection to the 
MIR source.  On the other hand, there were no multiple matching of the MIR sources for a given NIR 
source.  We summarize the results of  cross matching between the NIR and MIR bands in Fig. 16. 
Amongst the MIR bands, there were no duplicate matches, as presented in Table 3, since the number 
density of the MIR sources is much lower.

\subsubsection{Optical identification}

\begin{figure*}[ht!]
\centering
\resizebox{0.7\textwidth}{!}{\includegraphics{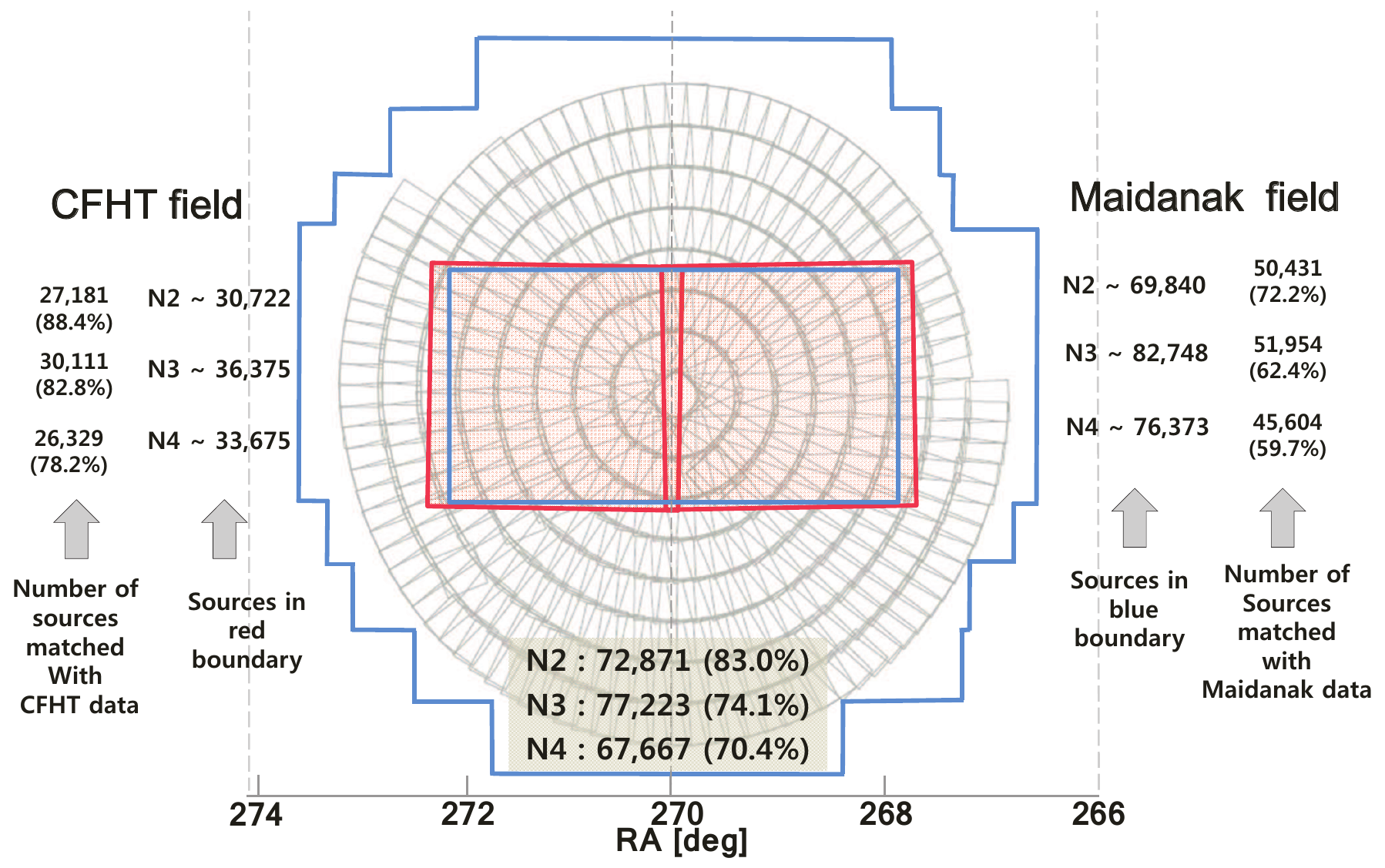}}
\caption{The results of the matching AKARI sources with the optical data covering two separate fields 
of the CFHT and Maidanak observations. On the left and right of the diagram, the results for the CFHT 
and Maidanak data are presented, respectively.  Overall results of the matching both sets of optical
data are given in the bottom of the figure.}
\label{fig 17}
\end{figure*}

On the left and right hand sides of Fig 17, the numbers of sources matched with each of the optical 
catalogs are presented. Note that the total number of  $N3$ sources is 104,170, while the number of 
them in the CHFT field is about 36,380.  Amongst these sources, 82.8$\%$ in the CFHT field have optical 
counterparts.  Similarly, the number of $N3$ sources located in the Maidanak field is about 82,750, 
which is more than twice the number of sources in the CFHT field, and 62.4$\%$ of these sources have 
Maidanak optical counterparts.  On the whole, 74.1$\%$ of the $N3$ sources have counterparts in either
the Maidanak or CFHT data.   As shown in the lower part of Fig. 17, 83.0$\%$ of the $N2$ and 70.4$\%$ 
of the $N4$ sources have optical counterparts. The rate of matching with optical data decreases with
NIR band wavelength. The matching rate  with the CFHT data is somewhat higher than with the Maidanak 
data, because the CFHT observations are deeper.

\begin{figure}[h!]
\centering
\resizebox{.9\hsize}{!}{\includegraphics{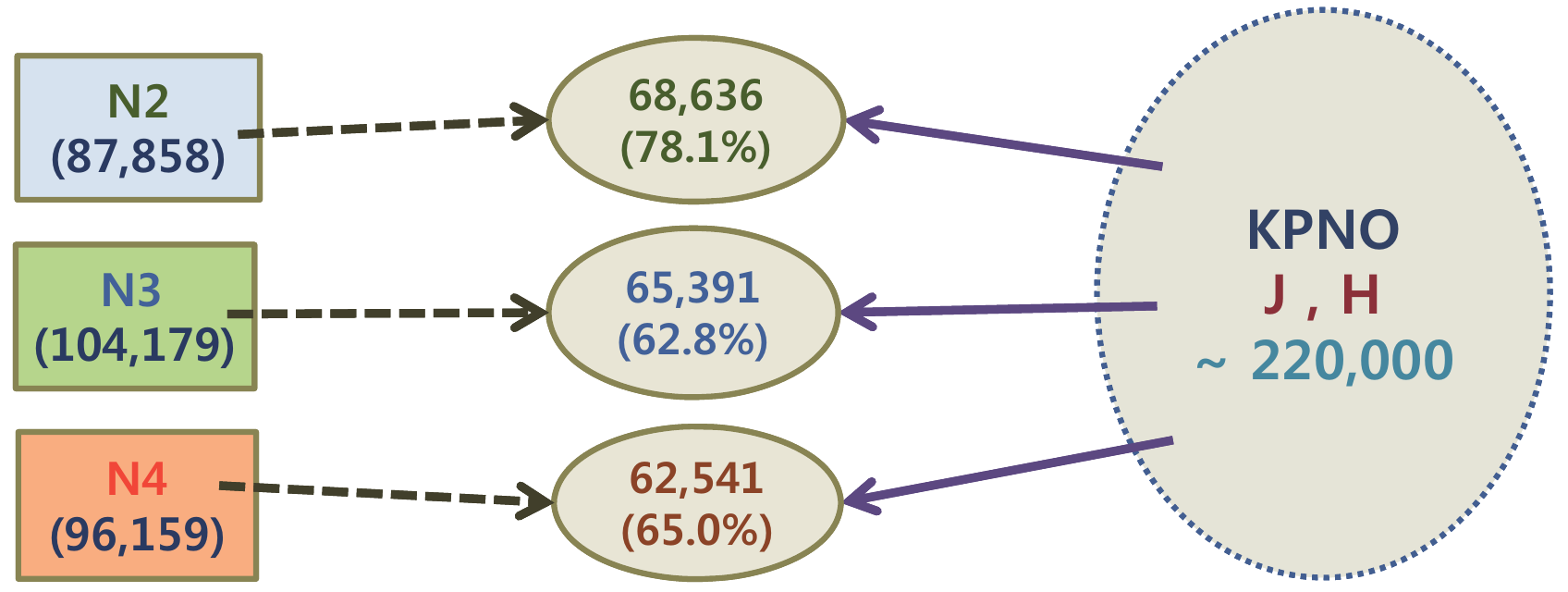}}
\caption{The number of sources in each of the NIR band detection catalogs matched with KPNO $J$, $H$
band data. About 78$\%$, 63$\%$, and 65$\%$ of the $N2$, $N3$, and $N4$ sources were found to have
counterparts in the $J$ and $H$ band data.}
\label{fig 18}
\end{figure}

We also examined the matching of the NIR sources with the KPNO $J$, $H$ data. The cross matching 
results are presented in Fig. 18. For the $N2$, $N3$ and $N4$ bands, 78 $\%$, 63 $\%$ and 65$\%$ have 
counterparts in KPNO data, respectively.

\section{CATALOGUE}

\subsection{Band-merging }

\begin{figure*}[t!]
\centering
\resizebox{0.6\hsize}{!}{\includegraphics{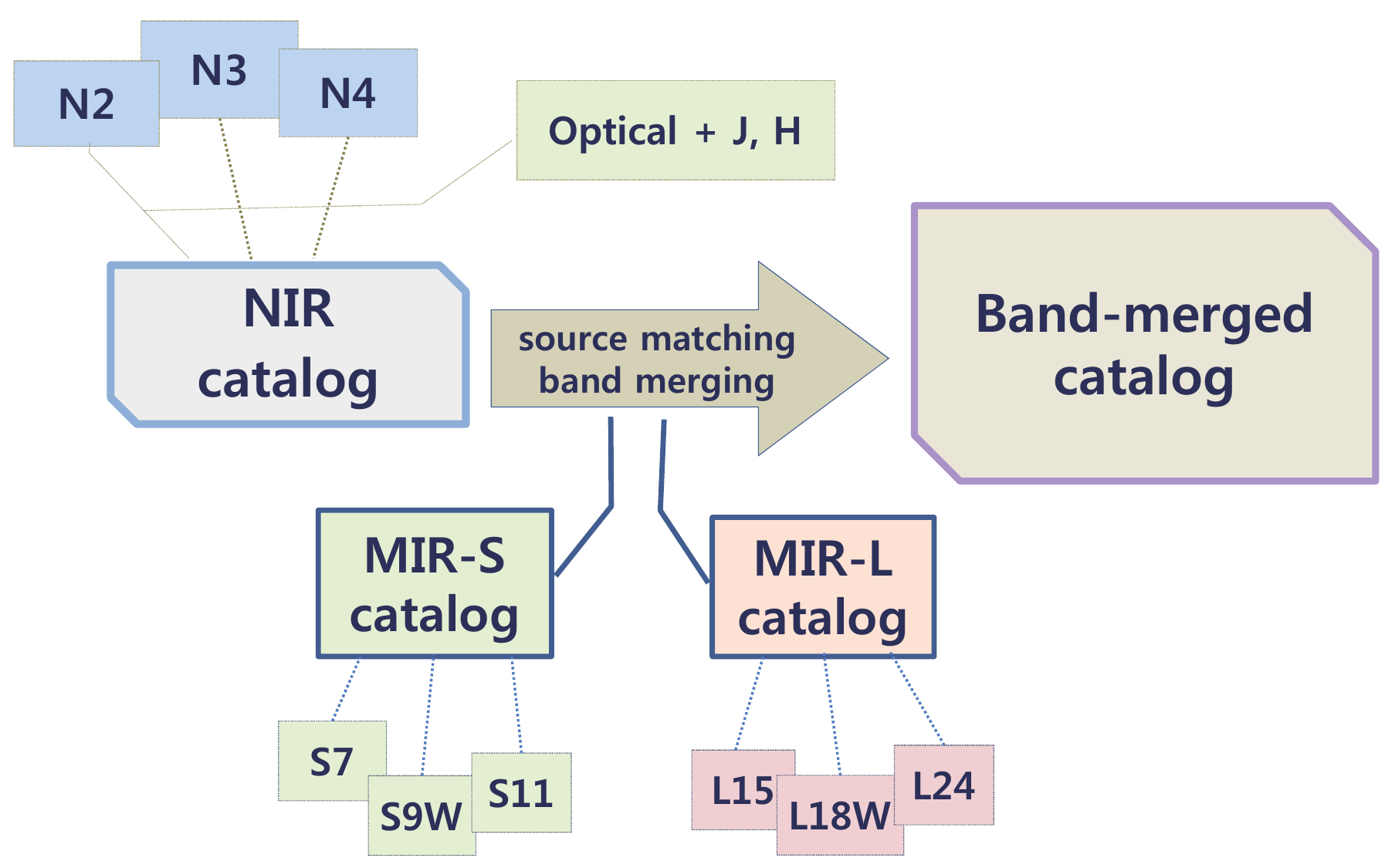}}
\caption{Schematic diagram describing the band merging procedure.  we first combined the three catalogs
of the NIR, MIR-S, and MIR-L channels. We then merged them into a band-merged catalog covering the
optical $u^{*}$ to the $L24$ band.  During this procedure, we checked the source IDs carefully in
order to avoid any duplication of the same entries. }
 \label{fig 20}
\end{figure*}

`
After the source matching  described above, we generated separate catalogs for the nine IRC bands. At 
this point, each catalog contained the photometric information from the other bands.  We merged those
catalogs step by step. As shown in Fig. 19, We made a NIR band catalog from the $N2$, $N3$, and $N4$ 
single-band catalogs by cross-matching the sources, and where possible, with the optical data.  The 
MIR-S and MIR-L catalogs were also generated with the same method as used for the NIR catalog. The 
final band-merging was based on the NIR catalog, and the sources in the MIR-S and MIR-L catalogs were 
compiled by matching of the entries using corresponding source IDs. For registered sources in the 
catalog, if there was no detection in a certain band, we assigned the dummy value 99.000. The NIR 
sources having no counterpart in any other bands are excluded to avoid the false objects caused by 
various artifacts. As explained in \S4.2, however, we carried out a careful eye inspection of the 
individual images of the MIR sources with no counterparts before this exclusion.

\subsection{Catalog format}

\begin{deluxetable}{rccccccccccccccccccccccccccccccccc}
\tablenum{4a}
\tabletypesize{\tiny}
\setlength{\tabcolsep}{0.03in}
\rotate
\tablecaption{NEP-Wide Infrared Point Source Catalogue\label{cat}\tablenotemark{a}}
\tablewidth{0pt}
\tablehead{
\colhead{ID}&\colhead{RA}&\colhead{Dec}& \multicolumn{3}{c}{- N2 -}& \multicolumn{3}{c}{- N3 -} &\multicolumn{3}{c}{- N4 -} &\multicolumn{3}{c}{- S7 -}& \multicolumn{3}{c}{- S9W -}& \multicolumn{3}{c}{- S11 -}& \multicolumn{3}{c}{- L15 -}& \multicolumn{3}{c}{- L18W -} \\
\colhead{num}&\colhead{[deg]}&\colhead{[deg]}&\colhead{[mag]} & \colhead{[err]}&\colhead{[flg]}&\colhead{[mag]}&\colhead{[err]}&\colhead{[flg]}&\colhead{[mag]}&\colhead{[err]}&\colhead{[flg]}& \colhead{[mag]} & \colhead{[err]}&\colhead{[flg]}&
    \colhead{[mag]} &\colhead{[err]} &\colhead{[flg]} &\colhead{[mag]} &\colhead{[err]} &\colhead{[flg]} & \colhead{[mag]} &\colhead{[err]} &\colhead{[flg]}& \colhead{[mag]} &\colhead{[err]} &\colhead{[flg]} \\
    \colhead{(1)}&\colhead{(2)}&\colhead{(3)}  & \colhead{(4)} &\colhead{(5)} &\colhead{(6)} &\colhead{(7)}&\colhead{(8)}&\colhead{(9)}&\colhead{(10)}&\colhead{(11)}&\colhead{(12)}&\colhead{(13)}&\colhead{(14)}&\colhead{(15)}&
    \colhead{(16)}&\colhead{(17)}&\colhead{(18)}&\colhead{(19)}&\colhead{(20)}&\colhead{(21)}&\colhead{(22)}&\colhead{(23)}&\colhead{(24)}&\colhead{(25)}&\colhead{(26)}&\colhead{(27)} &}
\startdata
     3 & 266.59579 & 66.47150 & 19.928 &  0.066 &  0 & 99.000 & 99.000 & -1 & 99.000 & 99.000 & -1 & 99.000 & 99.000 & -1 & 99.000 & 99.000 & -1 & 99.000 & 99.000 & -1 & 99.000 & 99.000 & -1 & 99.000 & 99.000 & -1  \\
     7 & 266.59860 & 66.55494 & 99.000 & 99.000 & -1 & 17.690 &  0.013 &  0 & 99.000 & 99.000 & -1 & 18.095 &  0.073 &  0 & 18.500 &  0.118 &  0 & 19.031 &  0.164 &  0 & 99.000 & 99.000 & -1 & 99.000 & 99.000 & -1  \\
    18 & 266.60712 & 66.53683 & 19.594 &  0.051 &  0 & 19.073 &  0.035 &  2 & 19.713 &  0.058 &  0 & 99.000 & 99.000 & -1 & 99.000 & 99.000 & -1 & 18.755 &  0.145 &  0 & 99.000 & 99.000 & -1 & 99.000 & 99.000 & -1  \\
   296 & 266.73248 & 66.44641 & 19.057 &  0.037 &  3 & 19.440 &  0.044 &  2 & 18.832 &  0.029 &  2 & 18.218 &  0.065 &  0 & 99.000 & 99.000 & -1 & 17.346 &  0.067 &  0 & 99.000 & 99.000 & -1 & 99.000 & 99.000 & -1  \\
   520 & 266.79562 & 66.14382 & 17.934 &  0.020 &  0 & 18.088 &  0.018 &  2 & 18.502 &  0.023 &  0 & 18.697 &  0.081 &  0 & 16.887 &  0.043 &  0 & 17.081 &  0.042 &  0 & 17.563 &  0.098 &  0 & 17.385 &  0.067 &  0  \\
  9509 & 267.66681 & 65.92631 & 14.171 &  0.003 &  0 & 14.694 &  0.003 &  0 & 15.335 &  0.003 &  0 & 16.319 &  0.019 &  0 & 16.029 &  0.022 &  0 & 17.089 &  0.044 &  0 & 99.000 & 99.000 & -1 & 99.000 & 99.000 & -1  \\
 24706 & 268.54489 & 67.36521 & 17.518 &  0.015 &  0 & 17.931 &  0.017 &  0 & 18.331 &  0.021 &  0 & 17.333 &  0.042 &  0 & 15.877 &  0.021 &  0 & 16.252 &  0.024 &  0 & 16.555 &  0.040 &  0 & 16.685 &  0.038 &  0  \\
\enddata
\tablefoot{
 $-$ Band merged catalog of AKARI/NEP-Wide sources. Col. (1): ID number. Col. (2)-(3): Coordinates, R.A.
 and Dec. based on N2 astrometry.  Col. (4)-(30): the IRC photometric information of the sources. For 
each band,  magnitudes, magnitude errors and flags are presented.
}
\tablenotetext{a}{This table contains only a subset of NEP-Wide sources. The complete version of the 
catalogs is in the electronic edition of the Journal. }
\label{nepcat}
\end{deluxetable}

\begin{deluxetable}{rccccccccccccccccccccccccc}
\tabletypesize{\tiny}
\tablenum{4b}
\setlength{\tabcolsep}{0.03in}
\rotate
\tablecaption{NEP-Wide Infrared Point Source Catalog\label{cat}\tablenotemark{b} - continued from the previous table}
\tablewidth{0pt}
\tablehead{
\colhead{ID}& \multicolumn{3}{c}{- L24 -} &\colhead{N\_flg}&     \multicolumn{2}{c}{$u^{*}$}&     \multicolumn{2}{c}{$g^{\prime}$}&     \multicolumn{2}{c}{$r^{\prime}$}&     \multicolumn{2}{c}{$i^{\prime}$}&     \multicolumn{2}{c}{$z^{\prime}$}&    \multicolumn{2}{c}{$B$}&     \multicolumn{2}{c}{$R$}&    \multicolumn{2}{c}{$I$}&   \colhead{stell}& \colhead{\# of src}& \colhead{$\Delta\theta$}  \\
\colhead{num}&\colhead{[mag]}&\colhead{[err]}&\colhead{[flg]}  &\colhead{[flg]}   &  \colhead{[mag]}&\colhead{[err]}&  \colhead{[mag]}&\colhead{[err]}& \colhead{[mag]}&\colhead{[err]}& \colhead{[mag]}&\colhead{[err]}& \colhead{[mag]}&\colhead{[err]}&\colhead{[mag]}&\colhead{[err]}& \colhead{[mag]}&\colhead{[err]}& \colhead{[mag]}&\colhead{[err]}&\colhead{  }& \colhead{  }      \\
\colhead{   }&\colhead{(28)} &\colhead{(29)} & \colhead{(30)} &\colhead{(31)} & \colhead{(32)}&   \colhead{(33)}&  \colhead{(34)}& \colhead{(35)}&  \colhead{(36)}& \colhead{(37)}&  \colhead{(38)}& \colhead{(39)}&  \colhead{(40)}& \colhead{(41)}&  \colhead{(42)}& \colhead{(43)}& \colhead{(44)}& \colhead{(45)}& \colhead{(46)}& \colhead{(47)}&  \colhead{(48)}& \colhead{(49)}&    \colhead{(50)}  }
\startdata
      3 & 99.000 & 99.000 & -1  &  0 &  99.000 & 99.000 & 99.000 & 99.000 & 99.000 & 99.000 & 99.000 & 99.000 & 99.000 & 99.000  & 99.000 & 99.000 & 22.310 &  0.170 & 20.570 &  0.070 & 0.460 & 1 & 1.146  \\
      7 & 99.000 & 99.000 & -1  &  0 &  99.000 & 99.000 & 99.000 & 99.000 & 99.000 & 99.000 & 99.000 & 99.000 & 99.000 & 99.000  & 19.780 &  0.040 & 17.730 &  0.040 & 16.550 &  0.020 & 0.980 & 1 & 2.053  \\
     18 & 99.000 & 99.000 & -1  &  0 &  99.000 & 99.000 & 99.000 & 99.000 & 99.000 & 99.000 & 99.000 & 99.000 & 99.000 & 99.000  & 99.000 & 99.000 & 21.470 &  0.160 & 20.570 &  0.150 & 0.180 & 1 & 0.341  \\
    296 & 99.000 & 99.000 & -1  &  0 &  99.000 & 99.000 & 99.000 & 99.000 & 99.000 & 99.000 & 99.000 & 99.000 & 99.000 & 99.000  & 99.000 & 99.000 & 99.100 & 99.000 & 20.850 &  0.120 & 0.010 & 1 & 1.072  \\
    520 & 99.000 & 99.000 & -1  &  0 &  99.000 & 99.000 & 99.000 & 99.000 & 99.000 & 99.000 & 99.000 & 99.000 & 99.000 & 99.000  & 99.000 & 99.000 & 99.100 & 99.000 & 21.360 &  0.110 & 0.190 & 1 & 2.552  \\
   9509 & 99.000 & 99.000 & -1  &  0 &  99.000 & 99.000 & 99.000 & 99.000 & 99.000 & 99.000 & 99.000 & 99.000 & 99.000 & 99.000  & 14.230 &  0.030 & 13.940 &  0.010 & 13.440 &  0.030 & 1.000 & 1 & 0.515  \\
  24706 & 16.661 &  0.079 & -1  &  0 &  99.000 & 99.000 & 99.000 & 99.000 & 99.000 & 99.000 & 99.000 & 99.000 & 99.000 & 99.000  & 19.800 &  0.050 & 18.650 &  0.040 & 18.270 &  0.030 & 0.030 & 1 & 0.592  \\
\enddata
\tablefoot{
 Col. (32)-(41): CFHT Megacam u$^{*}$, g$^{\prime}$, r$^{\prime}$, i$^{\prime}$, and z$^{\prime}$ 
magnitudes from Hwang et al. (2007) that are matched with AKARI/NEP-Wide sources.  Col. (42)-(47): 
Maidanak B, R, and I magnitudes from Jeon et al. (2010).  Col. (48): stellarity parameter from two 
optical data.  Col. (49): the number of optical counterparts within3$\arcsec$.  Col. (50): the angular 
separation of the optical counterpart from the position of AKARI source. }
\label{nepcat}
\tablenotetext{b}{This table is continued from the previous table and contains only a subset of NEP-Wide 
sources. The complete version of the catalogs is in the electronic edition of the Journal. }
\end{deluxetable}

The number of sources in the final catalog is about 114,800. In Table 4, an extract from the source 
catalog is shown for the purpose of illustration.  We give a short description below of the columns in 
the catalog.

 Column (1) contains the identification number of the sources.

 Columns (2) and (3) are the J2000 right ascension (R. A.) and the declination of a source in decimal 
degrees. The coordinates are based on the $N2$ astrometry. If there was no detection in the $N2$ image, 
the coordinates were based on the shortest wavelength identification.

 Columns (4), (7), (10), (13), (16), (19), (22), (25), and (28) are the  AB magnitudes of the sources 
in the AKARI bands from the $N2$ to $L24$ band.

 Columns (5), (8), (11), (14), (17), (20), (23), (26), and (29) provide the uncertainties in the 
magnitudes, which are the RMS errors in  the measured values.

 Columns (6), (9), (12), (15), (18), (21), (24), (27), and (30) are  the photometric flags from the 
SExtractor. 
    0  -  well isolated and clean image,
    1  - the source has neighbors, about 10$\%$ of the integrated area affected,
    2  - the object was blended with another one,
    3  - 1$+$2,
    4  - at least one pixel is saturated,
    8  - object is truncated, close to the image boundary,
   -1  - no photometry (not detected)

 Column (31) is the additional flag that indicates any caution about the use of the NIR photometry.
(this indicates partially damaged sources during the cosmic-ray rejection, masking of MUX-bleeding, 
and etc. )
    2 - for $N2$,  3 - for $N3$,  4 - for $N4$. \\

For the matching results with optical data, we present the magnitude, magnitude error, and stellarity 
information. In addition, the number of matched optical sources and the positional deviation of optical 
source from the AKARI WCS are included.

 Columns (32), (34), (36), (38), (40) are the  magnitudes in $u^{*}$, $g^{'}$, $r^{'}$, $i^{'}$, $z^{'}$ 
from the CFHT data.

 Columns (33), (35), (37), (39), (41) are the magnitude errors in $u^{*}$, $g^{'}$, $r^{'}$, $i^{'}$, 
$z^{'}$.

 Columns (42), (44), (46) are the  magnitudes in $B$, $R$, $I$ from Maidanak data.

 Columns (43), (45), (47) are the  magnitude errors in $B$, $R$, $I$.

 Column (48) contains stellarity information from optical data. Note that the stellarity values for the  
CFHT data were measured from the $g^{'}$, $r^{'}$, $i^{'}$, $z^{'}$ combined image (Hwang et al., 2007) 
and those of the Maidanak data were determined from the $R$- band image because of its highest S/N (Jeon 
et al., 2010). If a source is matched with both the CFHT and Maidanak data, we used the stellarity 
determined from the CFHT data.

Column (49) provides the number of optical sources matched to an AKARI source. For multiply matched 
sources, this is larger than 1.

Column (50) contains the distance between the astrometric coordinates of AKARI and optical data in 
arcsec. When an AKARI source is matched to optical data more than once (that is, column (49) $>$ 1), 
we chose the smallest number (the value of the closest one).

\section{NATURE OF THE SOURCES} 

The point sources in the NEP-Wide catalog are mostly either stars or galaxies.  In high-resolution 
optical images, the galaxies appear as extended sources unless they are very far away or very compact, 
whereas stars appear as point sources whose images follow the point spread function of the instruments. 
However, in the AKARI image, distant galaxies cannot be easily distinguished from those of stars 
because of the relatively large PSFs. In many cases, the nature of the sources can be identified by 
their spectral energy distributions (SEDs).

In contrast to the infrared images, the ground-based optical images have much smaller PSFs, allowing 
an easy distinction between the point and extended sources. In particular, the optical data for the 
NEP-Wide were all taken in  excellent seeing conditions. Both the CFHT and Maidanak data have typical 
PSF FWHMs smaller than one arcsecond.  The stellarity parameter given by SExtractor can thus be used 
to distinguish between point and extended sources, as shown in Fig. 20. The sources with stellarity 
parameters close to 1 are point-like sources, while those with values close to 0 are extended sources. 
Although the stellarity parameters are spread over all the possible values between 0 and 1, there is a 
clear dichotomy, unless the sources become too faint. For the following discussion, we selected the 
high-stellarity sources using an optical stellarity parameter ($> 0.8$) and an $r^{'}$ band magnitude 
cut ($< 19$) and designated them as `star-like' objects. 

\begin{figure}[h]
\begin{center}
\resizebox{\hsize}{!}{\includegraphics{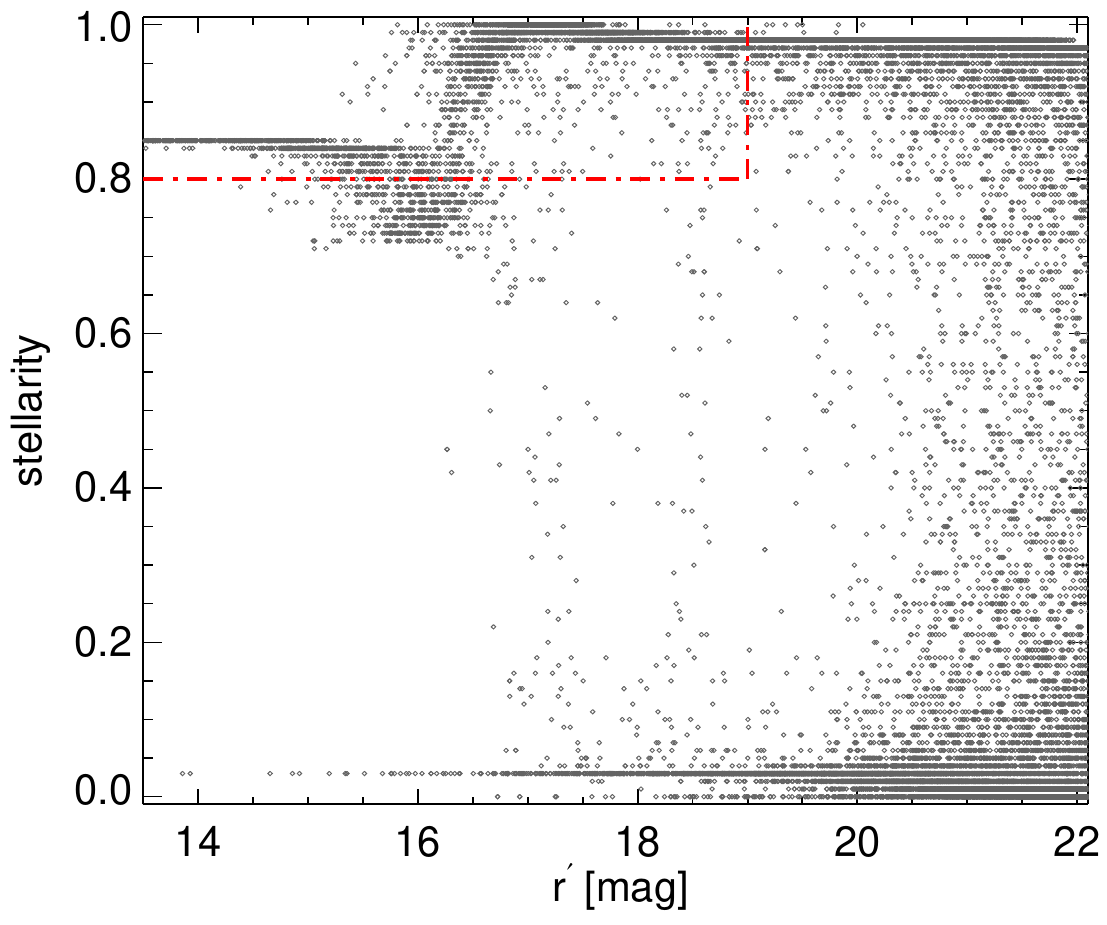}}
\caption{The stellarity parameter as a function of $r^{'}$ band magnitude derived from the CFHT data.
It can be clearly seen that the stellarity parameters are mostly either close to 1 or 0 unless the 
sources are very faint. Note that the sources brighter than 16 mag have relatively small 
stellarities because the central parts of their images are saturated.  
}
\label{fig 20}
\end{center}
\end{figure}

Stellar SEDs are usually determined by the surface temperature and the atmospheric metal abundances. 
Since the stellar surface temperature is usually higher than 3000K, the infrared parts of stellar SEDs 
can be approximated by the Rayleigh-Jeans spectrum.  Thus, we expect the stars to make a smaller 
contribution to the SED as the wavelength increases. However, infrared properties can often be modified 
by the presence of circumstellar material. Some stars with large amounts of circumstellar material
can be bright even in some MIR bands.

\begin{figure*}[ht!]
\centering
\resizebox{\hsize}{!}{\includegraphics{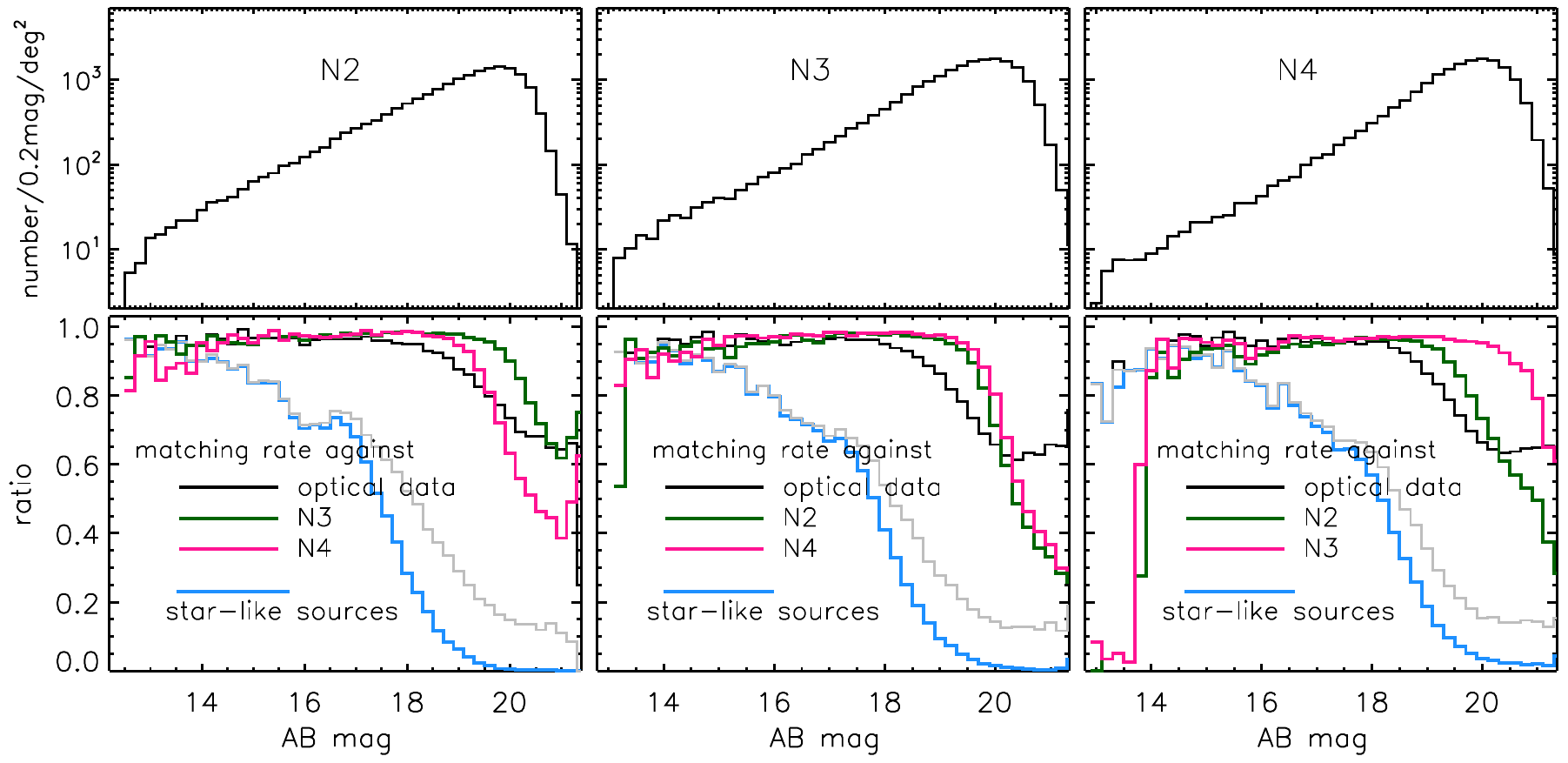}}
\caption{The source counts of NIR bands and the matching ratio against optical and other NIR bands.
Top panels show the distribution of sources as a function of magnitude per square degree per 0.2
magnitude bin. In the lower panels, the matching ratio against those in other NIR bands and optical
data are presented together.}
\label{fig 21}
\end{figure*}

Galaxies emit significant amounts of radiation in the infrared. Their SEDs strongly depend on galactic 
type and star formation rate (e.g., Polletta et al. 2007).  Since our catalog covers a wide range of 
wavelengths, the SED fitting of individual galaxies could provide useful information about their 
natures. However, this is beyond the scope of the present paper. Here we provide only general and 
statistical comments about the compositions of the sources in the present paper.

\subsection{Number counts and the source matching ratio}

\begin{figure*}[ht!]
\centering
\resizebox{\hsize}{!}{\includegraphics{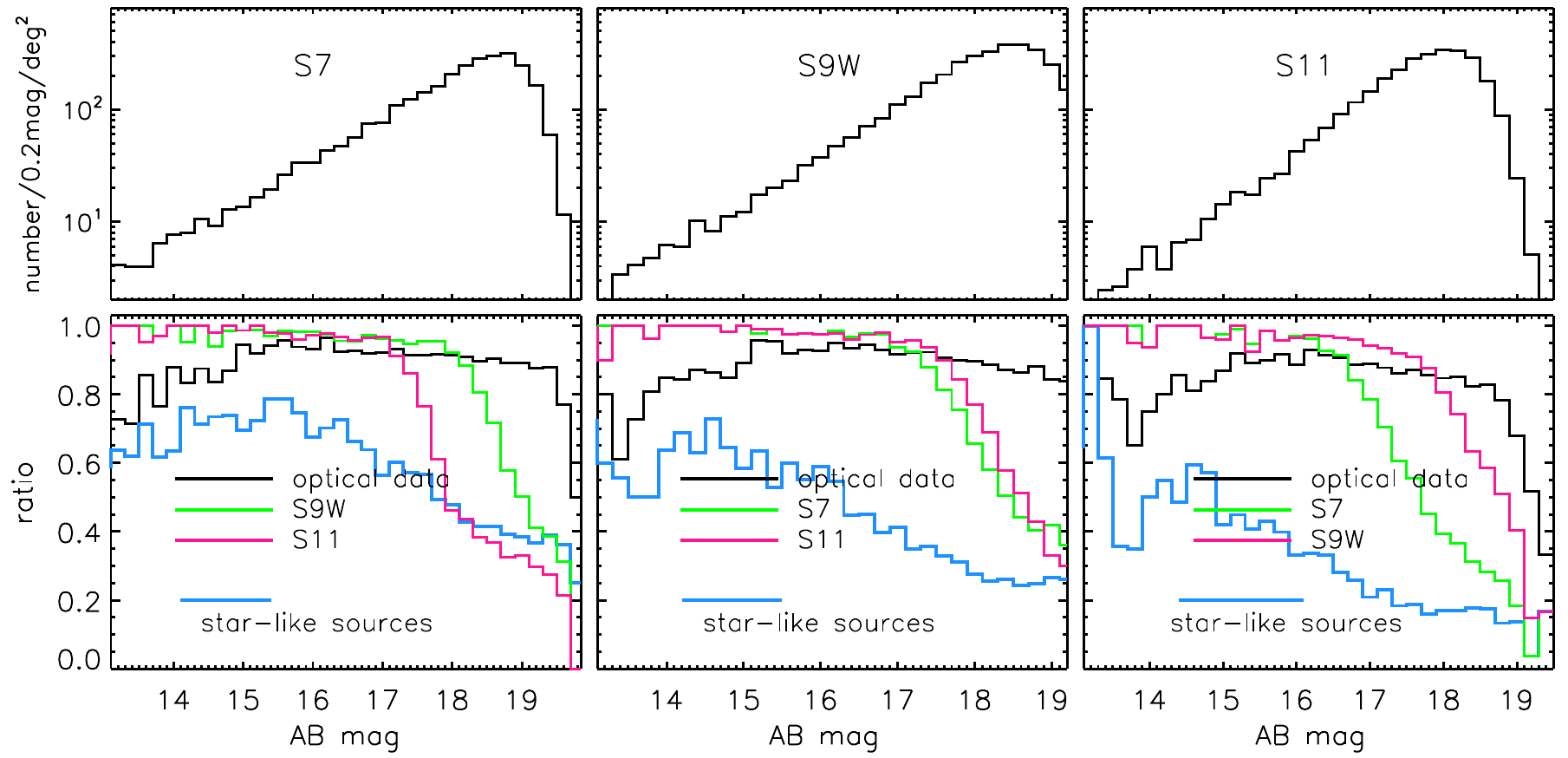}}
\caption{Same as Fig. 21, but for the MIR-S band sources.
The top panel shows the number of sources per square degree per 0.2 mag bin
as a function of magnitude. In the lower panel, the matching rates of the sources
against those in the other MIR-S bands and the optical data are presented together with the
fraction of star-like sources.}
\label{fig 22}
\end{figure*}

\begin{figure*}[ht!]
\centering
\resizebox{.9\hsize}{!}{\includegraphics{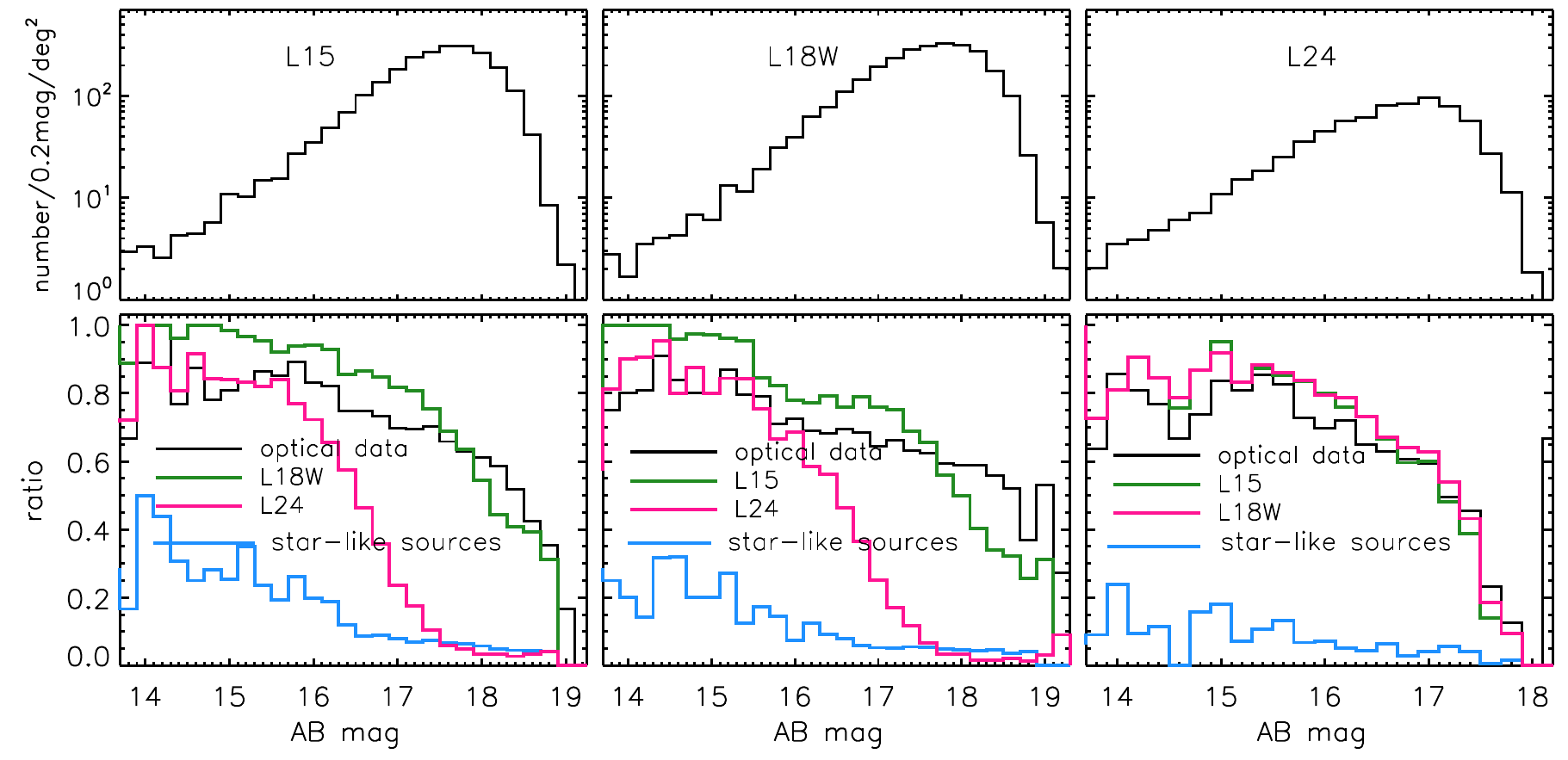}}
\caption{Same as Fig. 22, but for the MIR-L band sources.}
\label{fig 23}
\end{figure*}

We present number counts for each band and the cross-matching rate of the sources against optical data 
and the other bands as a function of magnitude. Fig. 21 shows the results for the NIR bands. The top 
panels show the number counts of NIR band sources which are confirmed through the cross-checking process 
with the other bands. The $N2$ sources brighter than 12.6 mag were excluded during the data reduction 
process because they were heavily saturated and generated MUXbleeds.  We found that the sources with 
no counterparts in the other data are mostly fainter than 19 mag in the $N2$ band. On the other hand, 
most of the false objects at bright magnitudes were found  to be residual streaks around bright stars 
through the visual inspection of the image. On the other hand, at fainter magnitudes, most of the false 
objects close to the detection limits are due to noises, because the comparison of common areas of the
NEP-Deep and NEP-Wide data shows that many of these faint sources do not appear as real sources in the
NEP-Deep,  whose detection limits are more than 1 mag deeper. 

In the lower panel, we showed the matching ratios against the optical, and the other two NIR bands 
data separately. The fraction of sources for which the stellarity is greater than 0.8 without any 
constraint on magnitude are indicated by the gray line in this panel. The fraction of star-like objects 
that are defined as those with the optical stellarity greater than 0.8 and $r$ magnitude brighter than 
19 are shown as blue dotted lines. Using the optical-IR SEDs of these objects, we confirmed that most 
of the star-like objects closely follow  black body spectra, suggesting that they are indeed likely to 
be stars.   Fig. 21 thus suggests that the sources brighter than 14 mag in the $N2$ band are mostly 
stars, even though the actual number of them is small. The fraction of high-stellarity objects rapidly 
decreases around 17 mag, and completely falls off to zero toward $\sim$ 20 mag. The features in the 
$N3$ and $N4$ band are generally similar to those in the $N2$ band. The properties of these objects in 
color-color diagrams are discussed in the following subsection. The overall fraction of star-like 
objects among the entire $N2$ sources is  about 17$\%$.  In the $N3$ and $N4$ bands, these fractions
are  about 15$\%$  and 16$\%$, respectively.

The same analysis was conducted for the MIR bands and the results are shown in Fig. 22 (MIR-S) and 
23 (MIR-L). The numbers of sources detected in the MIR bands are much smaller than those in the NIR 
bands.  Compared to the fraction of high-stellarity sources in the NIR bands, their fractions in the
MIR bands are small at all magnitudes, and decreases significantly toward the longer wavelength bands. 
However, the number of these sources neither rapidly decline nor become zero in the fainter regions.

Overall, about 22$\%$ of the entire NIR and 20$\%$ of the MIR sources do not have optical counterparts
within a 3$^{''}$ radius, while there are multiple optical counterparts for some sources. This multiple 
matching was most serious between the optical and MIR band sources. We found multiple optical 
counterparts for about 17$\%$ of the MIR sources during the MIR band merging. In these cases, we tried 
to identify the proper counterparts using the optical sources that had NIR counterparts from AKARI or 
in the $J$, $H$ bands. Through this process, we were able to assign the reliable counterparts with 
confidence. Despite this, the nearest neighbor does not seem to be the most likely optical counterpart 
for about ($\sim 3\%$) of the MIR sources. In those cases, we visually inspected the images, although 
this provided little help.  In the final catalog, all of the optical counterparts are indeed those 
found to be the nearest neighbors.

\subsection{Color-color and color-magnitude diagrams }

\begin{figure*}[ht!]
\begin{center}
\centering
\resizebox{.85\hsize}{!}{\includegraphics{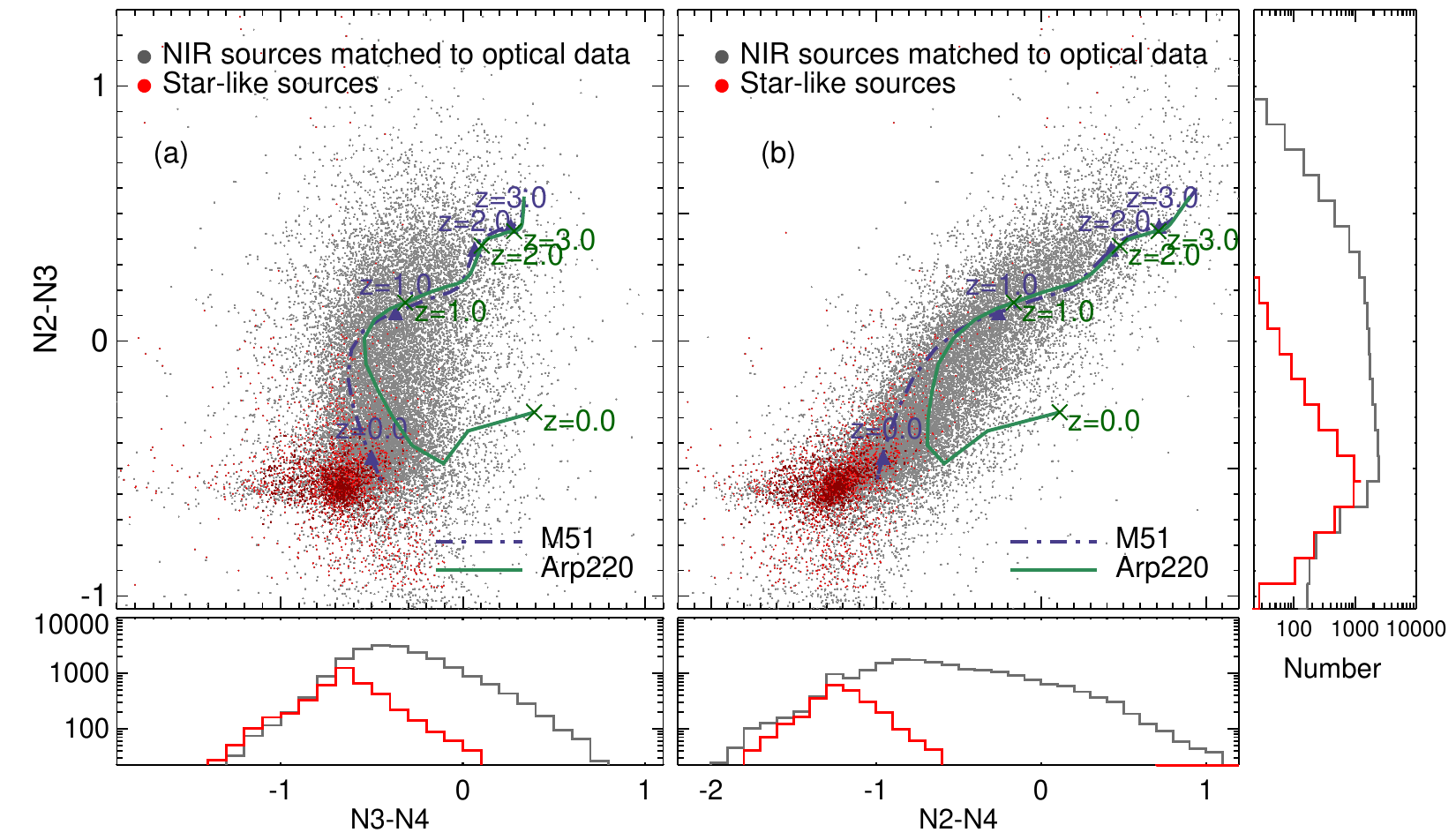}}
\caption{The color-color diagrams of the NEP-Wide NIR sources matching with optical data. (a) 
$(N2-N3)$ vs. $(N3-N4)$ color and (b) $(N2-N3)$ vs. $(N2-N4)$ color. Dark dots represent all of the 
sources having optical counterparts and red dots represent the `star-like' sources defined as having
the criteria of stellarity $> 0.8$ and $r^{'}$  $< 19$.  The histograms in both the right and lower 
panels show the number of sources per 0.2 magnitude bin.
\label{fig 24}}
\end{center}
\end{figure*}

\begin{figure*}[ht!]
\begin{center}
\centering
\resizebox{0.85\hsize}{!}{\includegraphics{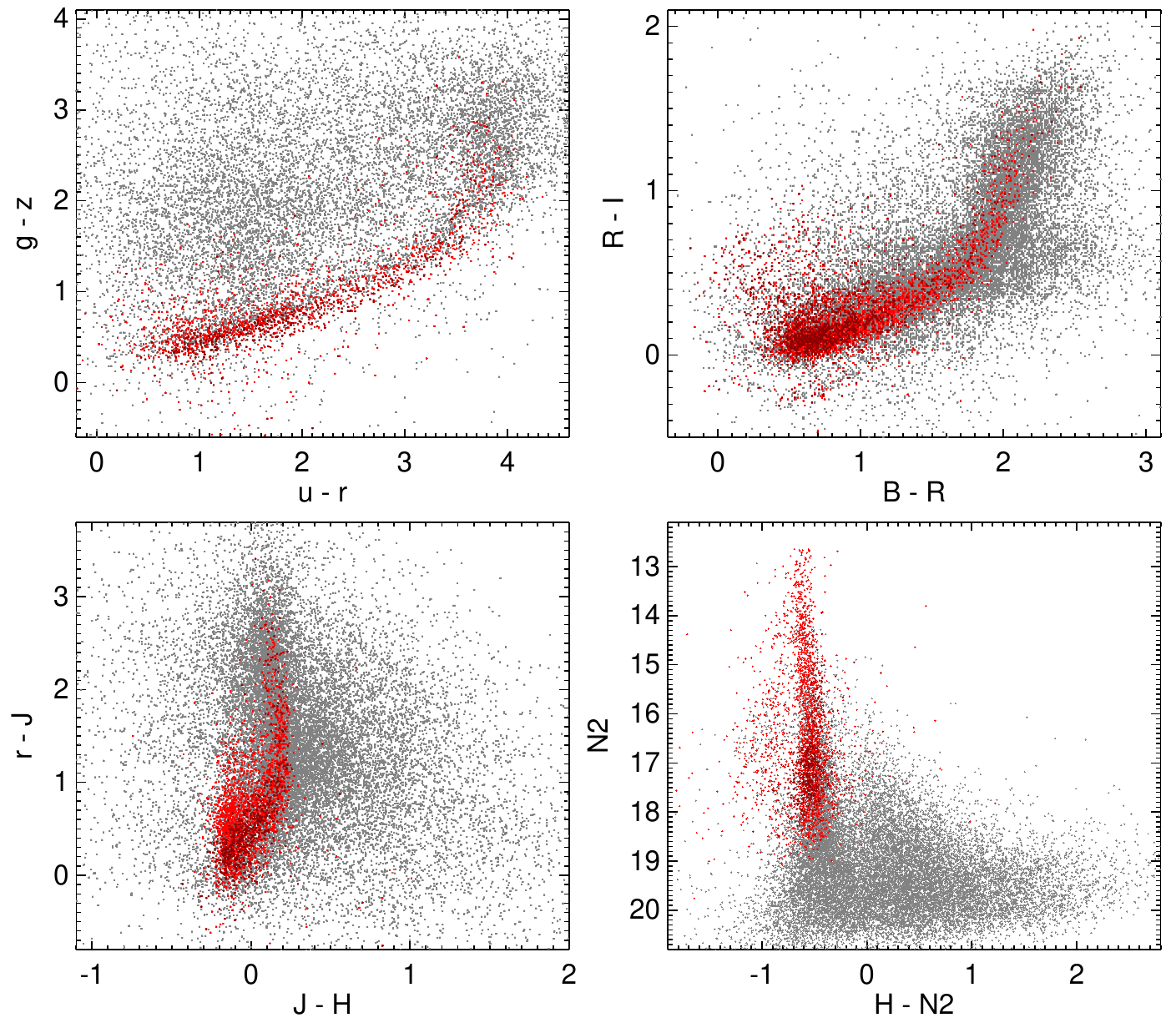}}
\caption{The color-color diagrams  and color-magnitude diagram of the NEP-Wide sources matched with 
the optical data. They show the various colors by optical bands from the CFHT and Maidanak and the 
near-IR $J$, $H$ bands.  Red dots represent the `star-like' sources defined as having  stellarity 
$> 0.8$ and $r^{'}$  $< 19$.
\label{fig 25}}
\end{center}
\end{figure*}

It is not easy to clearly establish the nature of the sources with the AKARI data alone. We attempted
to distinguish different types of sources using various color-magnitude and color-color diagrams. 

The star-like sources are plotted with red colors in each color-color diagram from Fig. 24 to 27.
The color-color diagrams for the three NIR bands of AKARI are shown in Fig. 24. The majority of
the star-like sources lie in the narrow range of colors of  $-0.7 < (N2-N3) < -0.4 $, $-1.5 < 
(N2-N4) < -1.0$ and $-0.9 < N3-N4 < -0.4$. The sources with stellarity greater than 0.95, magnitudes
in the range $14.5 < r <17.5$,  and identified as stellar sources by inspection of their images and
SEDs are overplotted in dark red.  We also present various optical-NIR color-color diagrams in
Fig. 25.
The star-like sources form a tight sequence in the ($g^{'}-z^{'}$) versus ($u^{'}-r^{'}$) color-color 
diagram, being located along the lower right edge. On the other hand, extended sources are more widely 
spread, occupying the entire region above the stellar-source sequence. In the ($R-I$) versus ($B-R$) 
color-color diagrams, however, the stellar sequence is somewhat broad and runs across the broader 
region occupied by the extended sources. The NIR colors such as $(J-H)$ and $(H-N2)$ are also useful 
in distinguishing the stars from galaxies (extended sources) as shown in the lower panels of Fig. 25.  
The ($J-N2$) color is not shown here but it is similar to the ($H-N2$) color.  The $N2$ vs. $(H-N2)$ 
color-magnitude diagram (CMD) in the last panel shows that the star-like sources are the brightest ones 
in the $N2$ band. By comparing these properties in color-color plots, we can separate stars effectively,
even though we are unable to identify all stars individually. The diagrams show that the high-stellarity 
sources are statistically good indicator for the selection of the stars although the criterion does
not perfectly identify stars in all cases.

\begin{figure*}[ht!]
\begin{center}
\centering
\resizebox{0.9\hsize}{!}{\includegraphics{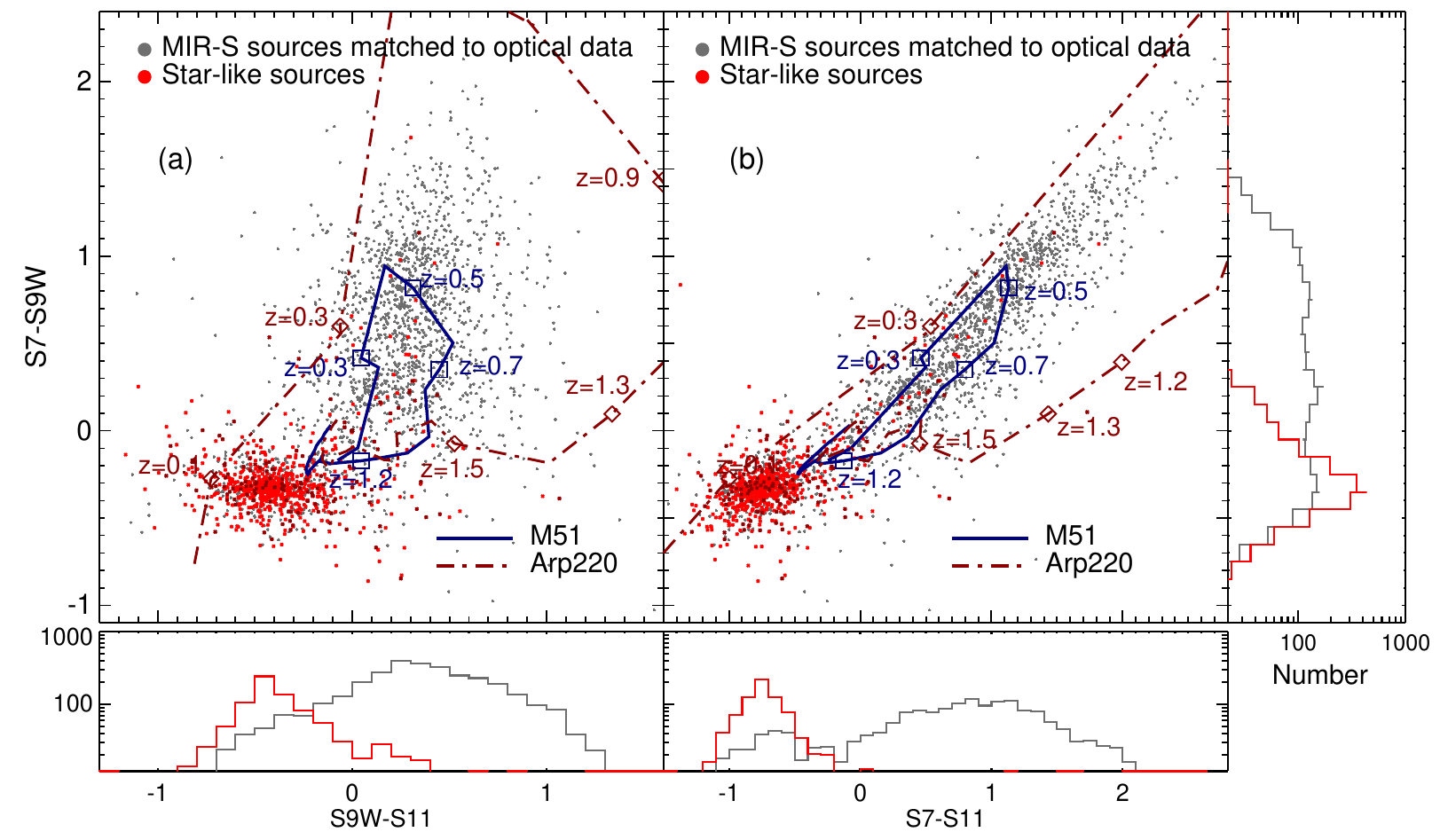}}
\caption{The color-color diagrams of the NEP-Wide MIR source matching with optical data. (a) $(S7-S9W)$
vs. $(S9W-S11)$ and (b) $(S7-S9W)$ vs. $(S7-S11)$ color-color diagrams.  All of the MIR sources having
optical counterparts are presented using dark dots in the diagrams Among them, the star-like sources
defined as those with stellarity greater than 0.8 are indicated by the red histogram.
\label{fig 26}}
\end{center}
\end{figure*}

\begin{figure*}[ht!]
\begin{center}
\resizebox{.9\hsize}{!}{\includegraphics{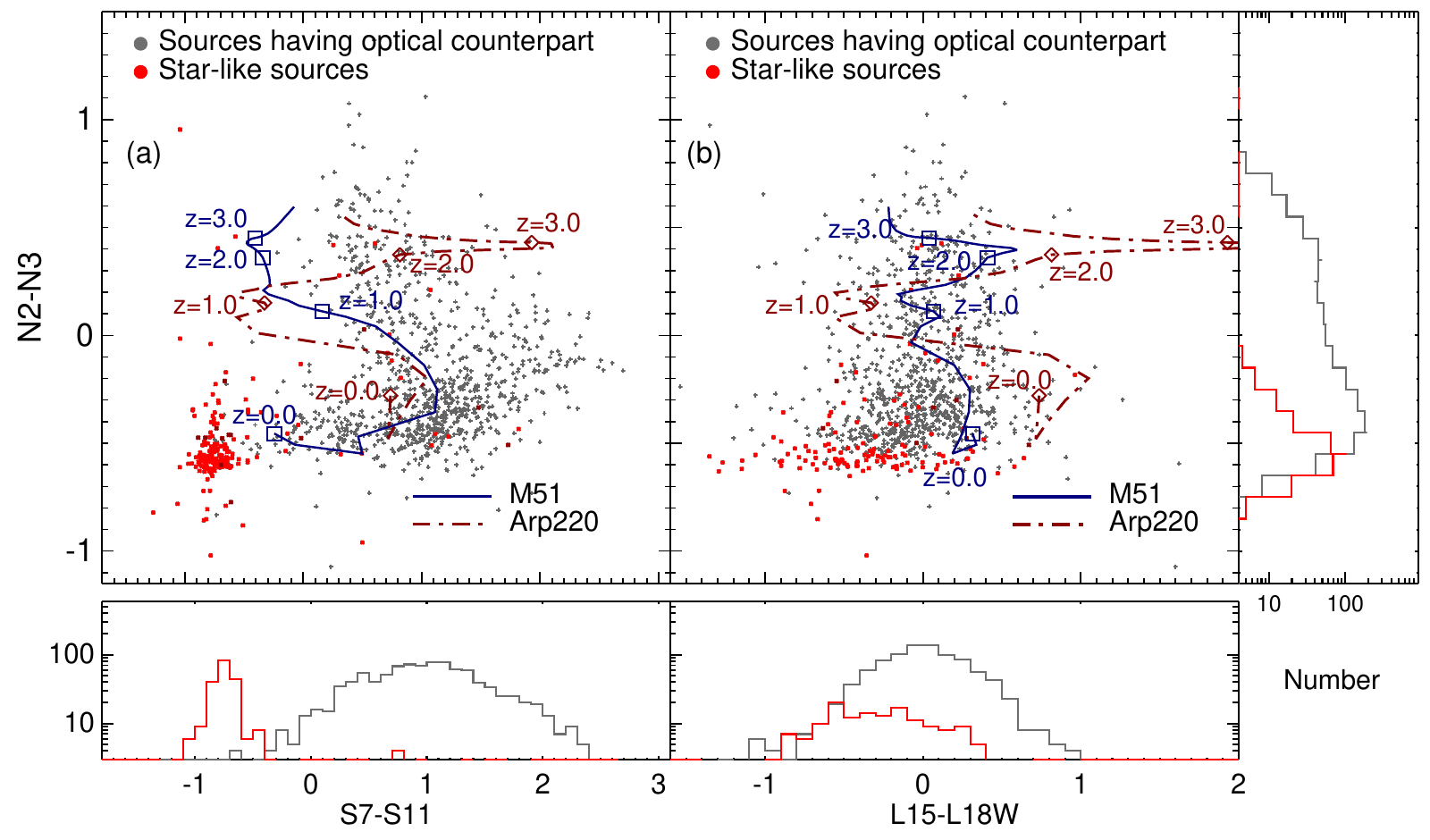}}
\caption{(a) $(N2-N3)$ vs. $(S7-S11)$, and (b) $(N2-N3)$ vs. $(L15-L18W)$ color-color diagrams. The
NEP-Wide  sources whose optical stellarities are known by cross-matching with the optical CFHT and
Maidanak catalogs are presented using dark dots in the diagrams.  Among them, the star-like sources
are plotted in red.
\label{fig 27}}
\end{center}
\end{figure*}

The colors of galaxies provide us with valuable information about their composition and history
(Fukugita et al. 1995). However, unlike point sources, extended sources show widely spread across 
the color-color diagrams as shown in Fig. 24 -- 27.  In Fig. 24, we have plotted the redshift tracks of
the templates (Silva et al. 1998) of the star-forming galaxy M51 and the typical ultra-luminous 
infrared galaxy (ULIRG) Arp220 over the diagrams.  Compared to these tracks, we find that the NEP-Wide 
sources are located close to the redshift sequence, although the scatter is quite large. This could 
mean that the galaxy SEDs are quite diverse, but in addition, a large fractions of the NEP-Wide 
sources are star forming galaxies at different redshifts. About 70\% of the MIR-selected sources in 
AKARI's early data were identified as star forming galaxies through the detailed inspection of their
infrared SEDs (Lee et al. 2007). Since the NIR parts of the SEDs of star-forming galaxies are dominated 
by late-type stars and thus rather homogeneous, the NIR colors are mostly determined by the galaxy 
redshifts.  The histograms show the distribution of the source density in each 0.1 magnitude bin. We 
also note that many sources are located quite far away from the star forming galaxy sequence. Some of 
these sources with very red NIR colors are likely to be active galactic nuclei (AGNs) (Lee et al. 2007).

In Fig. 26 and 27, we show the diagrams using the colors for the MIR bands, for which, the NEP-Wide
sources appear to have a much wider distribution. As for the sources used in the previous color-color
diagrams, optically matched sources are presented with gray colors and the star-like sources are
plotted with red colors. As shown in these figures, the MIR-S color-color diagrams $(S7-S9W)$,
$(S7-S11)$ and $(S9W-S11)$ show that the star-like sources are well segregated from the other sources,
but this feature gradually disappears in the longer wavelength band colors. Most of the stars (and the
early type galaxies) detected in the NIR bands seem to fade out in the MIR bands. Unlike the NIR and
MIR-S band colors, the MIR-L band colors do not seem to be helpful in classifying stars since the
MIR-L band color-color diagrams do not show any prominent features. The variation in the NIR colors are
mainly due to the wide range of galaxy redshift, while the MIR colors are sensitive to the star
formation rates, causing the wide spread in the MIR color-color diagrams. The number of sources detected 
in all IRC bands is about 1,000 and most of them seem to be late-type star forming galaxies. Many of 
these are identified as disk galaxies by the visual inspection of the optical images, and from their 
optical--IR SEDs. They are not located in a particular region and occupy a large area in the color-color
diagrams. However, various interesting sources seem to be included in the all-band detected sources.
Bright sources in the MIR-L bands that have radio counterparts are likely to be AGNs (Lee et al., 2009).
We found that many of them show a power law distribution of SEDs while some show PAH bumps (Takagi et
al., 2010).  Dozens of sources are bright ($< 17$ mag) in either the $L18W$ or $L24$ band and have
faint optical ($r^{'}$ or $R$) counterparts, seemingly suspected to be dust obscured galaxies (DOGs),
a class of high-redshift ULIRGs.  There are no sources with high-stellarity associated with faint MIR-L
sources.  In addition, the  MIR-L sources detected in only one band, having no counterparts in our other
data sets seem possibly to be very red objects.

\section{SUMMARY AND CONCLUSION}

We have carried out the reduction and analysis of the NEP-Wide survey data obtained by the AKARI/IRC. 
In order to reduce spurious detection, we masked out the regions affected by instrumental effects such 
as MUXbleeding trails, especially in the NIR bands. The detected sources were compared with the available
data at other wavelengths, including optical, ground-based near-IR observations, in addition to the 
other bands of AKARI. The areal coverage is about a 5.4 deg$^2$ circular field centered on the NEP. The 
5$\sigma$ detection limits of the survey are around 21 AB mag in the NIR bands, 19 -- 19.5 mag in the 
MIR-S bands, and 18.5 -- 18.8 mag in the MIR-L bands.

The ancillary optical data from the CFHT and the Maidanak observatory are sufficiently deep to identify 
most of the AKARI sources.  We carried out extensive comparisons by cross-matching of the sources among 
the photometric bands ranging from optical to mid-IR  wavelengths in order to confirm the validity of 
the detected sources and exclude the low-reliability sources. Using these results, we produced a 
band-merged source catalog covering the wavelength bands from the $u^{*}$ band to the MIR $L24$ band. 
This catalog contains about 114,800 entries. On the basis of this catalog, we have shown the 
characteristics of the sources using various color-color diagrams. By comparing and using optical 
stellarity, we found that the NIR and MIR-S band colors provide a reliable means of distinguishing stars
from galaxies. Except for the star-like objects, most of the NEP-Wide sources appear to be various types 
of star forming galaxy. The sources detected in all of the AKARI/IRC bands include interesting sources 
such as PAH galaxies, AGNs, ULIRGs, DOG candidates, or  MIR-bright early-type galaxies.

The NEP-Wide catalog covers a moderately large sky area with a wide wavelength range. It complements
the NEP-Deep catalogs of  Wada et al. (2008) and Takagi et al. (2012), which have better sensitivity and 
smaller angular coverage, but the same filter bands.

The Spitzer space telescope has also carried out large area surveys such as the Spitzer Wide-Area 
Infrared Extragalactic (SWIRE) Survey and First Look Survey (FLS). These surveys are carried out with 
all wide-band filters of Spitzer: 3.6 4.5, 5.8, and 8.0 $\mu$m with the Infrared Array Camera (IRAC) 
and 24, 70, and 160 $\mu$m with the Multiband Imaging Photometer for Spitzer (MIPS). This should be 
compared to the nearly continuous wavelength coverage of AKARI's NEP surveys from 2.4 to 24 $\mu$m. 
The FLS covered the area of about 5 deg$^2$, which is similar to that of the NEP-Wide. The 
survey area of SWIRE is about ten times larger, but composed of several different fields. 

This work is based on observations with AKARI, a JAXA project with the participation
of ESA, universities and companies in Japan, Korea, the UK, and Netherlands. This work 
contains many data obtained by the ground-based  Maidanak Observatory's 1.5 m, KPNO 2.1 m, and  
CFHT 3.5 m telescopes. This work was supported by the Korean Research 
Foundation grant 2006-341-C00018. AS and AP have been supported by the research grant of the
Polish National Science Centre N N203 51 29 38.

\end{document}